\documentclass[preprintnumbers,amssymb,prl,aps, notitlepage,twocolumn, superscriptaddress]{revtex4-2}

\usepackage{graphicx}
\usepackage{amsmath,amsbsy}
\usepackage{epsfig}
\newcommand{\comment}[1]{{}}
\newcommand{\pa}{\partial}
\newcommand{\en}{\nonumber\\}
\newcommand{\lcdm}{\ensuremath{\Lambda\mathrm{CDM}}}

 \renewcommand{\vec}[1]{\mathbf{#1}}
 
 \def\be{\begin{equation}}
\def\ee{\end{equation}}
\def\ba{\begin{eqnarray}}
\def\ea{\end{eqnarray}}
 
\begin{document}

\title{Symmetry of Cosmological Observables, a Mirror World Dark Sector, and the Hubble Constant}

\author{Francis-Yan Cyr-Racine}
\affiliation{Department of Physics and Astronomy, University of New Mexico, Albuquerque, New Mexico, USA 87106}
\author{Fei Ge}
\affiliation{Department of Physics and Astronomy, University of California, Davis, California, USA 95616}
\author{Lloyd Knox }
\affiliation{Department of Physics and Astronomy, University of California, Davis, California, USA 95616}

\email{fge@ucdavis.edu}

\date{\today}

\begin{abstract} 

We find that a uniform scaling of the gravitational free-fall rates and photon-electron scattering rate leaves most dimensionless cosmological observables nearly invariant. This result opens up a new approach to reconciling cosmic microwave background and large-scale structure observations with high values of the Hubble constant $H_0$: Find a cosmological model in which the scaling transformation can be realized without violating any measurements of quantities not protected by the symmetry. A ``mirror world" dark sector allows for effective scaling of the gravitational free-fall rates while respecting the measured mean photon density today. Further model building might bring consistency with the two constraints not yet satisfied: the inferred primordial abundances of deuterium and helium. 

\end{abstract}


\maketitle

{\it Introduction.}---{Different methodologies for determining the current rate of expansion, the Hubble constant $H_0$, are leading to discrepant results.
The most precise of the cosmological-model-dependent methods uses the {\it Planck} measurements of the cosmic microwave background (CMB). Assuming Lambda cold dark matter \lcdm\ the result is $H_0 = (67.49 \pm 0.53)$ km/sec/Mpc \cite{SPT:2021slg,Planck:2018vyg,aiola:2020}. The most precise of the more direct methods that are relatively independent of cosmological model assumptions comes from the SH$_0$ES team \cite{Riess:2011,Riess:2016jrr,Riess2019,riess2021}.
Using Cepheid-calibrated supernovae they find $H_0 = (73.04 \pm 1.04)$ km/sec/Mpc \cite[][hereafter R22]{Riess:2021jrx}, a 5$\sigma$ difference from the above result. 

The preference of \lcdm-dependent methods for a low $H_0$ is remarkably robust to choice of datasets and details of applications; it remains for both the ``inverse distance-ladder'' methods  \cite{percival:2010baryon, heavens:2014standard, aubourg:2015cosmological, cuesta:2015calibrating, Bernal:2016gxb, verde:2017length, Lemos:2019} that make minimal use of CMB data, and methods that rely on large-scale structure observations instead of CMB data \cite{DES:2017txv, DAmico:2019fhj, Ivanov:2019pdj, Colas:2019ret, Philcox:2020vvt,Zhang:2021yna}. 

Likewise, various averages of the more direct methodologies  \cite{freedman2012,Suyu:2016qxx,Birrer:2018vtm, Wong:2019kwg,Huang:2019yhh, Kourkchi:2020iyz, Reid:2019tiq, Freedman:2020dne, Freedman:2021ahq, Pesce:2020xfe, Khetan:2020hmh, Blakeslee:2021rqi,Birrer:2020tax}, including those that exclude Cepheid-calibrated supernovae, also lead to $ >4\sigma$ discrepancies with the CMB- and \lcdm-inferred value of $H_0$ \cite{DiValentino:2020vnx}.
On the other hand, there is no convergence of opinion yet among those using supernovae to measure $H_0$. Most notably, the Carnegie-Chicago Hubble Program \cite{Beaton:2016nsw,Hatt:2017rxl,Hatt:2018opj,Hatt:2018zfv,CSP:2018rag,Hoyt_2019,Beaton_2019,freedman2019,Jang_2021,Hoyt_2021} finds some inconsistencies in SH$_0$ES estimates of distances to nearby supernovae. For a review of the observational situation see Refs.~\cite{Freedman:2021ahq,Paul:2021abc}. }

A lot of recent theoretical work is inspired by the possibility that the tension arises from a failure of \lcdm.~For a wide-ranging discussion of possible avenues for a solution, see Ref.~\cite{knox20}, for a summary of many of the proposed models, see Ref.~\cite{divalentino21}, and for a ranking of their performance with respect to a common dataset, see Ref.~\cite{Schoneberg:2021qvd}. The search 
is difficult due to the variety of cosmological measurements, their sensitivity to the details of the cosmological models, their high precision, and their high degree of consistency with \lcdm. 

In this {\em Letter}, we provide new insight into the structure of cosmological models to help theorists navigate this challenging environment, as they search for solutions to the Hubble tension.\comment{, or otherwise look for alternative explanations for precision cosmological observations.} 
In particular, we present a transformation that leaves all distance ratios and the statistical properties of fractional maps of CMB temperature anisotropy (i.e., $\Delta T/T$), CMB polarization, and galaxy number overdensity invariant. This symmetry transformation, under which all relevant length and timescales in the problem are rescaled by a constant scaling factor $\lambda$ at all redshifts $z$ \footnote{Redshift is often used as a timelike variable in cosmology since the light that arrives here from a given event on our past light cone is stretched by the expansion by a factor of $1+z$, providing a monotonic relationship between time of emission and redshift $z$ in a continuously expanding universe.}, has its roots in the scale-free nature of primordial fluctuations.

A restricted version of this transformation was introduced in Ref.~\cite{Zahn_2003}, in which only the gravitational timescales $1/\sqrt{G \rho_i}$, where $\rho_i$ is the mean density of the $i$th component, were scaled. This leads to an approximate symmetry that is severely broken at small angular scales and in polarization {on all scales}. By extending their transformation to include a scaling of the photon scattering rate, we have found a symmetry of the above-mentioned observables that is exact in the limit of equilibrium recombination {and zero neutrino mass}. Crucially, we find that {symmetry-breaking effects of real-world departures from these limits are} mild.

We also report here on partial progress toward the use of this symmetry to solve the $H_0$ tension. Significant observational constraints prevent us from a straightforward scaling of gravitational rates by the scaling of mean densities. To evade these constraints, and still reap the benefits of the scaling transformation, we are led quite naturally to the addition of a mirror world dark sector.
Dark sector models that contain a mirror (or ``twin'') sector \footnote{Models have also been proposed that contain multiple copies of the SM, see e.g.~Refs.~\cite{Dvali:2007hz,Dvali:2009ne,Arkani-Hamed:2016rle}.} that has the same particle content and gauge interactions as the standard  model (SM), have been extensively studied in the literature (see e.g.~Refs.~\cite{Chacko:2005pe,Chacko:2005vw,Chacko:2005un,Barbieri:2005ri,Craig:2013fga,Craig:2015pha,GarciaGarcia:2015fol,Craig:2015xla,Farina:2015uea,Farina:2016ndq,Prilepina:2016rlq,Barbieri:2016zxn,Craig:2016lyx,Berger:2016vxi,Chacko:2016hvu,Csaki:2017spo,Chacko:2018vss,Elor:2018xku,Hochberg:2018vdo,Francis:2018xjd,Harigaya:2019shz,Ibe:2019ena,Dunsky:2019upk,Csaki:2019qgb,Koren:2019iuv,Terning:2019hgj,Johns:2020rtp,Roux:2020wkp,Ritter:2021hgu,Curtin:2021alk,Curtin:2021spx}) in the context of the little hierarchy problem. To scale the photon scattering rate, one can alter the primordial fraction of baryonic mass in helium $Y_{\rm P}$, which is the method we use here.
We show below that such models can, in principle, exploit the scaling symmetry to accommodate a higher value of $H_0$ with CMB observations.

A strict implementation of the scaling transformation, however, leads to a conflict between the measurements of $Y_{\rm P}$ and the primordial deuterium abundance with expectations from big bang nucleosynthesis (BBN) as well as the $Y_{\rm P}$ necessary for scaling up the scattering rate.
Thus, we have essentially remapped the problem that has proven to be quite difficult, of reconciling a high $H_0$ with phenomenologically complicated CMB data, to a problem of reconciling a high $H_0$ with the observationally inferred primordial abundances of helium and deuterium. Our work motivates the search for models that can solve this reframed problem \cite{Fields:2019pfx}. {Finally, although we focus here on inferences from CMB observations, {\em all} of the cosmological-model-dependent inferences of $H_0$ we referred to above, including those in \cite{percival:2010baryon,heavens:2014standard, aubourg:2015cosmological,cuesta:2015calibrating,Bernal:2016gxb,DES:2017txv,verde:2017length,Lemos:2019,DES:2017txv,DAmico:2019fhj, Ivanov:2019pdj,Colas:2019ret,Philcox:2020vvt,Zhang:2021yna}, would be impacted similarly; i.e., 
replacing \lcdm\ with a model that solves our reframed problem would also reconcile these inferences with a high $H_0$, due to the nature of the symmetry.}

{\it The scaling transformation.}---
Let us assume, for now, that recombination happens in equilibrium and neutrinos are massless.
Then the only length scales in the linear perturbation evolution equations in the \lcdm\ model, written with the scale factor $a = 1/(1+z)$ as the independent timelike variable, are the gravitational timescales of each of the $i=1$ to $N$ components $1/\sqrt{G\rho_i(a)}$ and the photon mean free path between electron scatters, $1/\left[\sigma_T n_e(a)\right]$. As a result, if we consider the linear evolution of a single Fourier mode with wave number $k$, any fractional perturbation, such as $\delta \rho_{\rm m}({\vec k},a)/\rho_{\rm m}(a)$ satisfying the evolution equations will also satisfy them when transformed by a uniform scaling of all relevant (inverse) length scales (including $k$) by some factor $\lambda$.
Since the initial conditions in \lcdm\ do not introduce a length scale (the spectrum of initial perturbations is a power law)%
, the statistical properties of fractional perturbations
are independent of $\lambda$ except for an overall amplitude. Dependence on the amplitude can be removed \cite{Zahn_2003} by extending the scaling transformation to include $A_s \rightarrow A_s/\lambda^{(n_{\rm s}-1)}$ where $A_{\rm s}$ is the amplitude of the primordial power spectrum at some fiducial value of $k$, and $n_{\rm s}$ is the spectral index of the power law power spectrum of initial density perturbations.

In Ref.~\cite{Zahn_2003}, this transformation was introduced but without the photon scattering-rate scaling, {which is critically important for our purposes}. Including it, the transformation leads to an exact symmetry in the limit of equilibrium recombination \footnote{We could drop this caveat by extending the transformation further so that the atomic reaction rates scale appropriately as well. Doing so would also extend the challenge (significantly) of finding a model for which a direction in the parameter space corresponds to the scaling transformation.}, massless neutrinos \footnote{The very mild symmetry breaking from non-zero neutrino masses could be removed by also scaling them by a factor of $\lambda$.}, and linearized equations \footnote{We include this caveat because we do not have a proof that the symmetry holds in the full non-linear theory. It might. We do know it holds to second order in perturbations; {\bf see our Supplementary Material, which includes \citep{Ma:1995ey, Zahn_2003, Bartolo:2007ax}}.} of distance ratios and the statistical properties of maps made in projection on the sky of quantities such as $\delta \rho/\rho$, and fractional CMB temperature and polarization anisotropies. These include galaxy clustering power spectra, shear power spectra, galaxy-shear cross-correlations, fractional CMB temperature and polarization power spectra, the temperature-polarization cross spectrum, and the CMB lensing spectrum. The invariance exists for the \lcdm\ model, and any other model as long as additional length scales (if any, such as {those} related to mean curvature or neutrino mass) are properly scaled as well. 
Absent the introduction of new length scales, the full transformation can be written as
\begin{equation} \label{eq:scalingtransform}
\begin{split}
\sqrt{G\rho_i(a)}  \rightarrow \lambda  \sqrt{G \rho_i (a)}, \ \ & \ \
 \sigma_{\rm T} n_e(a)  \rightarrow  \lambda \sigma_{\rm T}n_e(a)\\
  {\rm and} \ \ A_{\rm s}  \rightarrow  A_{\rm s}/\lambda^{(n_{\rm s}-1)}.
\end{split}
\end{equation}

{\it Symmetry breaking}.---The transformation given in Eq.~\eqref{eq:scalingtransform} is severely constrained by
observations that are sensitive to \emph{absolute} densities of cosmological components. Most importantly, we know very precisely the mean energy density of the CMB today from measurements by FIRAS of its flux density across a broad range of wavelengths \cite{Fixsen1996,fixsen09}.  
By anchoring $\rho_\gamma$, this measurement severely limits our ability to exploit the scaling transformation to raise $H_0$ \footnote{ A similar point was made recently in Ref.~\cite{ivanov2020h}}. 

Other important effects that break the above symmetry arise from departures from thermodynamic equilibrium, as emphasized in Ref.~\cite{Zahn_2003}. Unlike periods of equilibrium, during which we have no sensitivity to the rates of the reactions that are maintaining equilibrium, periods in which equilibrium is lost provide us with valuable sensitivity to the relevant reaction rates. If we then assume that such microphysical rates are known, we can gain sensitivity to the expansion rate. A prime example is BBN, where sensitivity of the yield of helium {and deuterium} to nuclear reaction rates allows one to infer, from measurements of {helium and/or deuterium}, the expansion rate during BBN, and thus, through the Friedmann equation, the mass or energy density of the Universe at that time.
Similarly, hydrogen recombination is an out-of-equilibrium process which is sensitive to atomic reaction rates, and 
thus breaks the symmetry of the Eq.~\eqref{eq:scalingtransform} transformation. We will see that the impact of this latter symmetry breaking on our parameter constraints is mild.

{\it A mirror world dark sector and free $Y_{\rm P}$}.---Zahn and Zaldarriaga \cite{Zahn_2003} did not provide a physical mechanism by which one could realize their transformation, other than by varying $G$, as their goal was only analytic understanding. We now introduce a physical mechanism that, while not allowing for the transformation as strictly written, permits a complete mimicry of its effects, so that the same invariance is achieved.

By extending the \lcdm\ model to include a dark copy of the photons, baryons, and neutrinos (see, e.g.,~Refs.~\cite{Blinnikov:1983gh,Ackerman:2008gi,Feng:2009mn,Agrawal:2016quu,Foot:2002iy,Foot:2003jt,Foot:2004pa,Foot:2004wz,Foot:2007iy,Foot:2011ve,Foot:2013vna,Foot:2014mia,Foot:2014uba,Foot:2016wvj,Ciarcelluti:2004ik,Ciarcelluti:2004ip,Ciarcelluti:2008vs,Ciarcelluti:2010zz,Ciarcelluti:2012zz,Ciarcelluti:2014scd,Cudell:2014wca}), all with the same mean density ratios as in the visible sector, we can effectively mimic the $\sqrt{G \rho_i}$ part of the scaling transformation while evading the constraint from FIRAS. The dark photons (which have temperature $T_{\rm D}$) are a replacement for the additional visible photons that would violate the FIRAS constraint. The dark baryons [implemented as ``atomic dark matter" (ADM) \cite{Goldberg:1986nk,Fargion:2005ep,Khlopov:2005ew,Khlopov:2008ty,Kaplan:2009de,Khlopov:2010pq,Kaplan:2011yj,Khlopov:2011tn,Behbahani:2010xa,Cline:2012is,Cyr-Racine:2013ab,Cline:2013pca,Cyr-Racine:2013fsa,Fan:2013tia,Fan:2013yva,McCullough:2013jma,Randall:2014kta,Khlopov:2014bia,Pearce:2015zca,Choquette:2015mca,Petraki:2014uza,Cirelli:2016rnw,Petraki:2016cnz,Curtin:2020tkm}] allow us to scale up the total baryonlike density without changing the well-constrained (visible sector) baryon-to-photon ratio. The dark neutrinos allow us to scale up the effective number of free-streaming neutrino species from its \lcdm\ value of $N_{\rm eff}^{\rm FS} = 3.046$ \footnote{We note that using the more recent estimate of $N_{\rm eff}^{\rm fs} = 3.044$ \cite{Froustey:2020mcq} would have minimal impact on our results.}, preserving the well-constrained ratio of free-streaming to tightly coupled relativistic particle densities \cite{Hou:2011ec,Follin:2015hya,Baumann:2015rya}.

In our implementation, the new mirror dark sector interacts purely gravitationally with the visible sector and the CDM. Therefore, we can mimic a \lcdm\ model with scaled-up densities if the perturbations in the mirror world evolve in the same way as what they replaced in the visible sector and thus provide the same source to the metric perturbations. For this to be the case, the dark photons must transition from tightly coupled to freely streaming when the visible photons do. We thus ensure that the ADM recombines at approximately the same time as regular hydrogen by keeping the ratio $B_{\rm D}/T_{\rm D}$ fixed, where $B_{\rm D}$ is the binding energy of the ADM, and by setting the dark fine structure constant and dark proton mass equal to those in the light sector. For simplicity we assume there are no dark versions of helium or heavier nuclei.
We can keep the Thomson scattering rate on the scaling trajectory by adjusting $Y_{\rm P}$. At fixed baryon density, we have $n_e(z) \propto x_e(z) (1-Y_{\rm P})$ where $x_e(z)$ is the fraction of free electrons. So to scale the scattering rate appropriately, approximating $x_e(z)$ as fixed, we send $(1-Y_{\rm P}) \rightarrow \lambda(1-Y_{\rm P})$.

\begin{figure}[t]
\includegraphics[trim={0 0cm 0 0cm},clip,width=\linewidth]{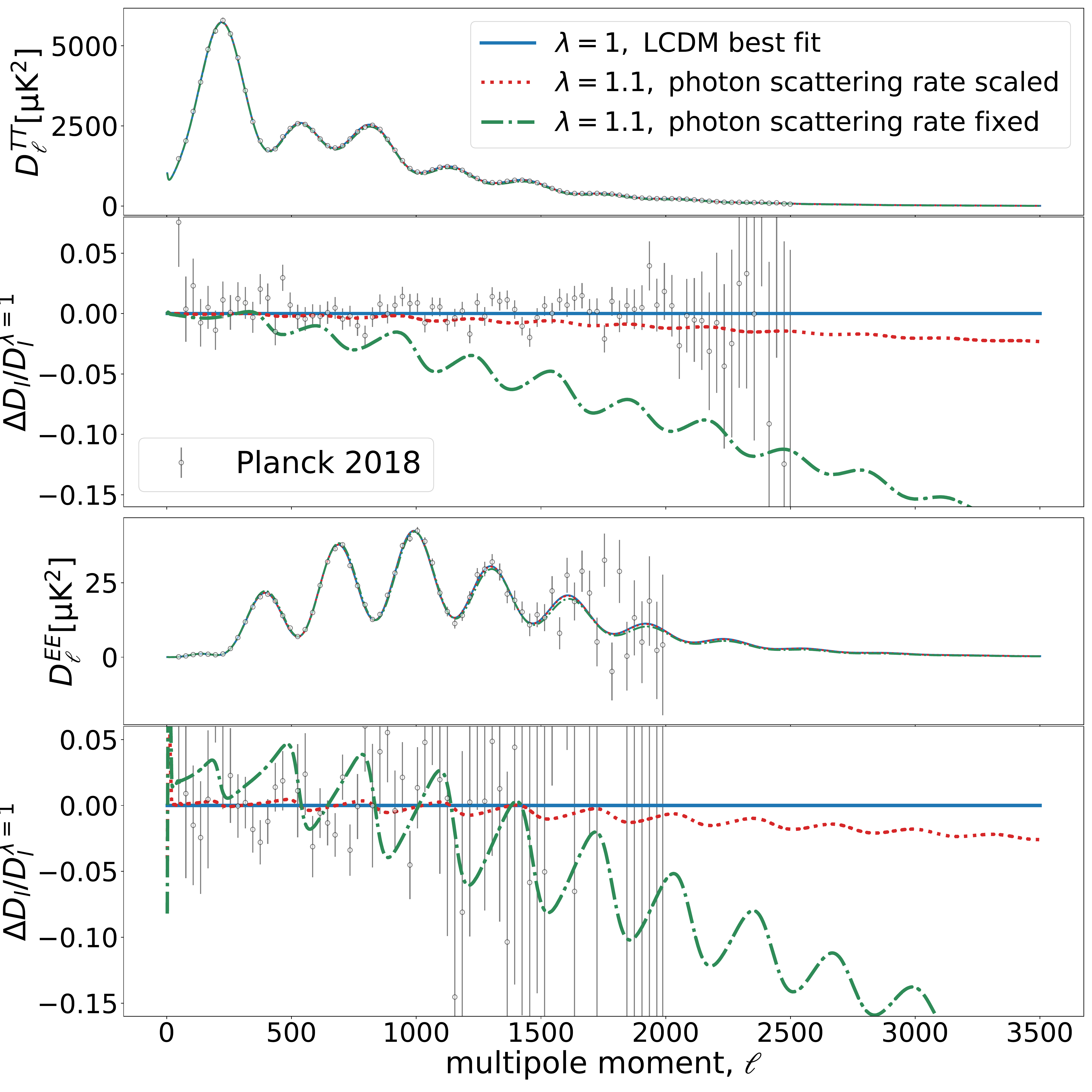}
\caption{CMB temperature (TT) and polarization (EE) power spectra with (red and green) and without (blue) scaling by $\lambda = 1.1$ from the best-fit  $\lcdm$ model for dataset 1. Both $\lambda = 1.1$ cases are scaled by use of the mirror world dark sector, with (red) and without (green) scaling of the photon scattering rate. Data points are from {\it Planck} 2018 \citep{Planck:2019nip}. Fractional differences with the best-fit \lcdm\ model are shown in the second and bottom panels.}
\label{fig1:scale}
\end{figure}

In Fig.~\ref{fig1:scale}, we show the temperature (TT) and polarization (EE) power spectra for the best-fit \lcdm\ model given dataset 1 (defined below) as well as spectra for the models scaled from that one with $\lambda = 1.1$, with and without the photon-scattering-rate scaling included. The scaled spectra with no photon scattering scaling differ significantly from the \lcdm\ spectra. The small differences between the fully scaled spectra and the \lcdm\ spectra are {primarily} due to the symmetry-breaking effects of the atomic reaction rates affecting hydrogen recombination.  

\begin{table}[t]
\label{tab:datasets}
\begin{tabular}{ | c | c|}
\hline
 Label & Dataset    \\
 \hline
 1 &  {\it Planck} TT, TE, EE, lensing + BAO \\
 \hline
 2 &  Dataset 1 plus R21 ($H_0 = 73.2 \pm 1.3$ km/s/Mpc)  \\
 \hline 
\end{tabular}
\caption{Definition of datasets 1 and 2. We use the Plik TT+TE+EE, Lowl$_{\rm T}$, Lowl$_{\rm E}$, and lensing {\it Planck} likelihoods described in Ref.~\cite{Planck:2019nip}. Baryon acoustic oscillations (BAO) datasets are 6dFGS \cite{Beutler:2011}, SDSS MGS \cite{Ross:2014qpa} and BOSS DR12 \cite{BOSS:2016wmc}. R21 is SH$_0$ES measurement \cite{riess2021}, of which R22 is a recent update.}
\end{table}

\begin{table}[t]
\label{tab:models}
\begin{tabular}{|c|c|}
\hline
 Label & Model space    \\
 \hline
 A & \lcdm\ + $\lambda$ (scaling enforced), $x_e(z)$ fixed  \\
 \hline
 B & \lcdm\ + $\lambda$ (scaling enforced), $x_e(z)$ calculated   \\
 \hline
 C & \lcdm\ + $T_{\rm D}$ + $ N_{\rm eff}^{\rm FS}$ + $f_{\rm ADM}$ + $Y_{\rm P}$ \\
 \hline 
\end{tabular}
\caption{Definition of model spaces A, B \& C. For model space A the one non-\lcdm\ parameter is $\lambda$. We restrict the additional components including $Y_{\rm P}$ to the scaling solution and (artificially) hold $x_e(z)$ fixed to its \lcdm\ best-fit value, for both the dark and light sectors. We scale $Y_{\rm P}$ from its BBN-consistent \lcdm\ dataset 1 best-fit value of 0.2454. Model space B differs only in that we calculate the visible sector $x_e(z)$ using the atomic reaction rates (and the code RECFAST \citep{Seager:1999bc, Wong:2007ym}) and the dark sector ionization evolution as in Ref.~\cite{Cyr-Racine:2013ab}. 
Model space C only differs from B in that we allow $Y_{\rm P}$, the effective number of free-streaming neutrinos $N_{\rm eff}^{\rm FS}$, and the fraction of mirror world (or ``atomic") dark matter $f_{\rm ADM}$ to depart from their scaling values. For all model spaces we adopt the uniform prior $1.00001< \lambda < 1.3$, set the ratio of dark to light photon temperatures to $T_D/T_\gamma=\left(\lambda^2-1\right)^{1/4}$, and, although it introduces a new length scale, we take one of the neutrino species to have a mass of 0.06 eV \cite{Esteban:2020cvm}. We modified CAMB \citep{Lewis:1999camb} to solve the relevant Einstein-Boltzmann equations and used CosmoMC \citep{Lewis:2002mc} to calculate parameter posterior densities. }
\end{table}

\begin{figure}[t]
\includegraphics[trim={0 0cm 0 0cm},clip,width=\linewidth]{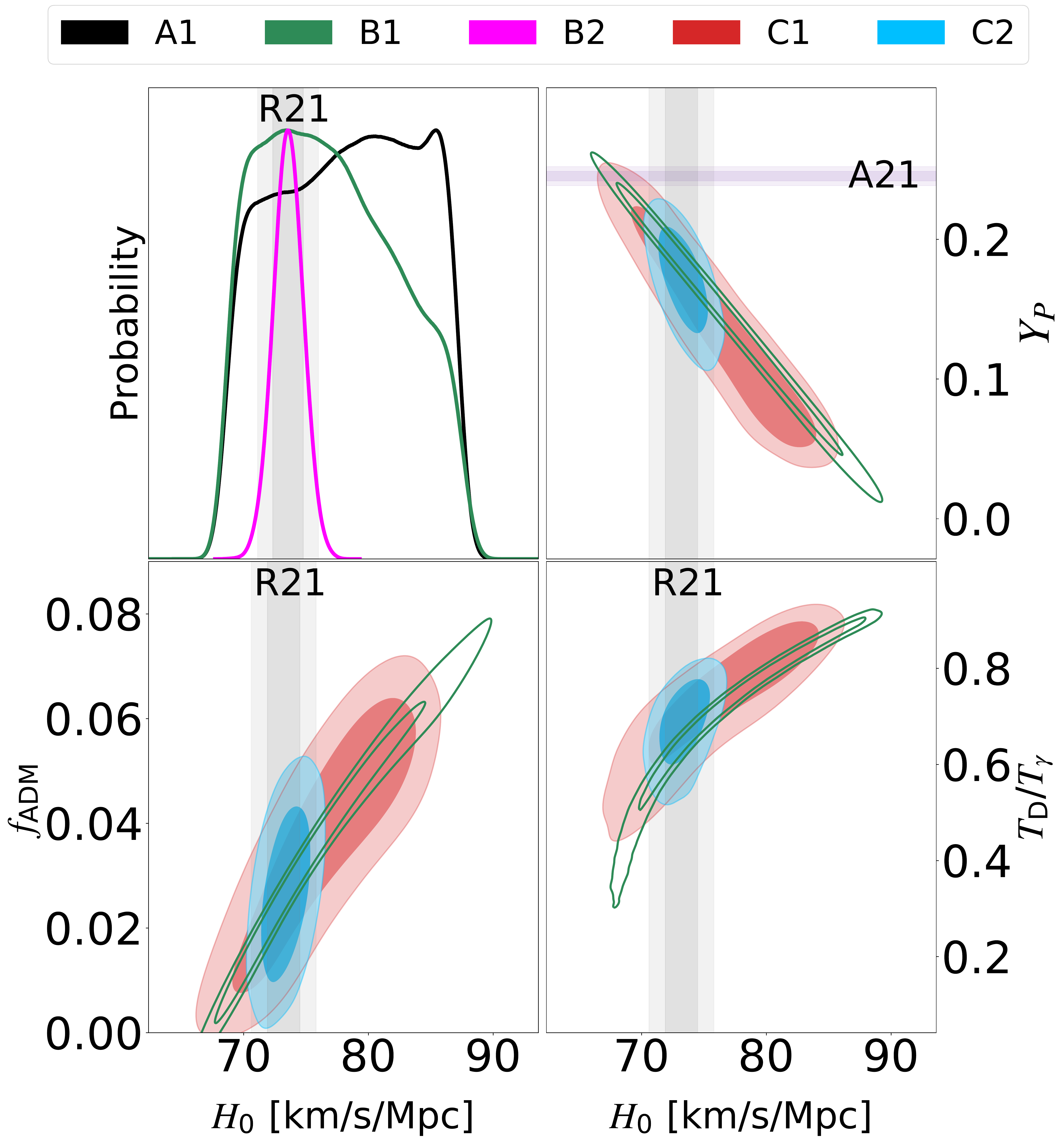}
\caption{Constraints on parameters from datasets 1 and 2 given models A, B and C (see Tables I and II). Top left panel: the (unnormalized) posterior probability density of $H_0$. Other panels: the 68\% and 95\% contours of equal probability density in the $H_0-Y_{\rm P}$, $H_0-f_{\rm ADM}$, and $H_0-T_{\rm D}/T_{\rm \gamma}$ planes. The gray band ``R21" shows the 1$\sigma$ and 2$\sigma$ constraint on $H_0$ from R21. The purple band ``A21" shows the same for $Y_{\rm P}$ from \citep[][hereafter A21]{aver2021}.}
\label{fig2:max_dof}
\end{figure}

{\it Results}.---We now explore the parameter constraints in the framework just described. We define datasets 1 and 2 and model spaces A, B, and C, in Tables I and II. The results are shown in Fig.~\ref{fig2:max_dof}. In the upper left panel
 we see the expected results from A1: The posterior is almost flat throughout the prior region \footnote{This prior is made necessary by our use of a mirror dark sector to mimic a $\Lambda$CDM model with scaled-up densities. It ensures that the energy densities of dark photons and dark atoms are always positive.}. 
 This is the numerical manifestation of the exact symmetry we have presented. 
 
 For B1, the $H_0$ posterior remains quite broad. The soft symmetry-breaking effects of nonequilibrium recombination, evident in Fig.~\ref{fig1:scale}, are sufficiently degenerate with variation of other \lcdm\ parameters to avoid significant constraint on $H_0$. The tension with R21 has been completely eliminated. The very slight preference for $H_0 = 73.2$ km/s/Mpc in B1 over the \lcdm\ value is a parameter-volume effect; the best-fit \lcdm\ model has a nearly identical $\chi^2$ value as the best-fit B model constrained to $\lambda = 1.08$, lower by $\Delta \chi^2 =$ 0.2.

In the other three panels of Fig.~\ref{fig2:max_dof} we show how well departures away from scaling are constrained by the data, if they are not prevented by fiat. The B1 contours lie over the constraints given by model C; as expected, the scaling direction is preferred by the data. We also see that the region of high posterior probability density extends to parameter values far from the scaling solution, an indication of some freedom that more detailed model building could exploit. For C2, we find that $f_{\rm ADM} = 0.027 \pm 0.011$ and $T_{\rm D} = (0.68 \pm 0.06)T_\gamma$. Meanwhile, the value of  $\sigma_8=0.808\pm0.011$ is nearly unchanged (if {not} slightly lower) from its \lcdm\ value, as expected.

In Fig.~\ref{fig2:max_dof} we also see a problem: The $Y_{\rm P}$ values consistent with R21 are inconsistent with inferences from spectral observations of hot ``metal-poor" gas such as the A21 finding of $Y_{\rm P} =  0.2453 \pm 0.0034$. From C2 we have $Y_{\rm P} = 0.170 \pm 0.025$, a 3.0$\sigma$ 
difference. Compounding this trouble, the additional light relics, if we do not otherwise alter the standard thermal history, would increase the BBN-expected $Y_{\rm P}$  \citep{Consiglio:2017pot}, in C2, to 
$0.2614 \pm 0.0038$.

{\it Discussion}.---We have found important constraints on the project of turning the scaling transformation into a solution to the $H_0$ tension. The first of these, from FIRAS, can be accommodated with a mirror world dark sector. Constraints from light element abundances lead us to {the} articulation of two additional targets for model building: (i) a new mechanism for increasing the photon scattering rate and (ii) additional model features that would bring BBN predictions for helium and deuterium abundances in line with observations. 

Item (i) follows from the fact that the $Y_{\rm P}$ required for consistency with R21 in models B and C is 3$\sigma$ too low compared to the inference from observations in A21, hence strongly suggesting that modifying $Y_{\rm P}$ is not a promising way to increase the photon scattering rate. Item (ii) follows since the extra relativistic species in the mirror sector required by R21 alter predictions of $Y_{\rm P}$ and deuterium. We have for C2 $\Delta N_{\rm eff} = 1.3 \pm 0.34$. On the other hand, BBN consistency with $Y_{\rm P}$ and deuterium measurements leads to $\Delta N_{\rm eff} = -0.17 \pm 0.28$ \cite{Fields:2019pfx}.  

 One idea worth exploring, to boost the photon scattering rate, is a heating of the baryons by a spectral distortion in the Wien tail at frequencies beyond FIRAS's reach \footnote{J. Chluba, private communication}. A time-varying electron mass \cite{Sekiguichi:2021, Hart:2020} is another possible solution that has had some phenomenological success \cite{Sekiguichi:2021}, and could even be well motivated by a supersymmetric gravity sector \cite{Burgess:2021qti,Burgess:2021obw}. Successfully fitting the light element abundances could be achieved, for instance, by reheating the dark sector in the post-BBN era (see, e.g.,~Refs.~\cite{Aloni:2021eaq}), or by possibly introducing interactions between the light and dark sectors that could affect the predicted yields; our inferred $T_D/T_\gamma$ even puts decoupling of the dark and light sectors in the right ballpark for this to occur.

 Our scenario requires a mirror sector that is phenomenologically close to the SM, albeit with a lower temperature. In its simplest realization, such a sector appears tightly constrained by particle collider data (see e.g.~Ref.~\cite{Burdman:2014zta}). Encouragingly however, preliminary work \cite{Blinov:2021mdk} indicates that such a mirror sector could be successfully built. 
  More generally, our work opens the possibility of relaxing cosmological constraints (see e.g.~Refs.~\cite{Craig:2016lyx,Chacko:2018vss,Dunsky:2019upk,Bansal:2021dfh}) on such mirror scenarios, given mechanisms to adjust the photon scattering rate and light-element abundances.

We note that having $\sim3\%$ of dark matter in ADM can lead to interesting astrophysical phenomenology such as exotic compact objects \cite{Shandera:2018xkn,Singh:2020wiq}, dark stars \cite{Curtin:2019lhm,Curtin:2019ngc}, exotic gamma-bright point sources \cite{Agrawal:2017pnb}, and dark disks within galaxies \cite{Fan:2013tia,Fan:2013yva,McCullough:2013jma,Randall:2014kta,Kramer:2016dew,Kramer:2016dqu,Schutz:2017tfp,Buch:2018qdr,Roux:2020wkp}.

{\it Conclusions}.---We have {generalized a scaling transformation so that it now} allows for large changes in the cosmological model while preserving the precisely measured and feature-rich CMB temperature and polarization spectra, as well as many other cosmological observables. Implementing this transformation, while evading constraints from FIRAS, leads us directly to a mirror world dark sector, a type of dark sector that has been proposed before for independent reasons.
 
 While the scaling symmetry does not, by itself, provide an end-to-end solution to the $H_0$ tension, we have used it to single out the photon scattering rate and light-element abundances as players with important roles in the discrepancy. We have thus provided clear model-building targets for the community to explore. 
 
\begin{acknowledgments}
{\it Acknowledgments:} F.G. and L.K. were partially supported by the U.S. Department of Energy Office of Science. F.-Y.~C.-R is supported by the National Science Foundation (NSF) under Grant No. AST-2008696. We thank B.~Fields, D.~Green, D.~Liu, and M.~Luty for useful conversations and the UNM Center for Advanced Research Computing, supported in part by the NSF, for providing some of the computing resources used in this work. Part of this work was performed at the Aspen Center for Physics, which is supported by NSF Grant No. PHY-1607611.
\end{acknowledgments}

\bibliographystyle{apsrev}
\bibliography{main}

\begin{thebibliography}{176}
\expandafter\ifx\csname natexlab\endcsname\relax\def\natexlab#1{#1}\fi
\expandafter\ifx\csname bibnamefont\endcsname\relax
  \def\bibnamefont#1{#1}\fi
\expandafter\ifx\csname bibfnamefont\endcsname\relax
  \def\bibfnamefont#1{#1}\fi
\expandafter\ifx\csname citenamefont\endcsname\relax
  \def\citenamefont#1{#1}\fi
\expandafter\ifx\csname url\endcsname\relax
  \def\url#1{\texttt{#1}}\fi
\expandafter\ifx\csname urlprefix\endcsname\relax\def\urlprefix{URL }\fi
\providecommand{\bibinfo}[2]{#2}
\providecommand{\eprint}[2][]{\url{#2}}

\bibitem[{\citenamefont{Balkenhol et~al.}(2021)}]{SPT:2021slg}
\bibinfo{author}{\bibfnamefont{L.}~\bibnamefont{Balkenhol}}
  \bibnamefont{et~al.} (\bibinfo{collaboration}{SPT}) (\bibinfo{year}{2021}),
  \eprint{2103.13618}.

\bibitem[{\citenamefont{Aghanim et~al.}(2020{\natexlab{a}})}]{Planck:2018vyg}
\bibinfo{author}{\bibfnamefont{N.}~\bibnamefont{Aghanim}} \bibnamefont{et~al.}
  (\bibinfo{collaboration}{Planck}), \bibinfo{journal}{Astron. Astrophys.}
  \textbf{\bibinfo{volume}{641}}, \bibinfo{pages}{A6}
  (\bibinfo{year}{2020}{\natexlab{a}}), \eprint{1807.06209}.

\bibitem[{\citenamefont{Aiola et~al.}(2020)\citenamefont{Aiola, Calabrese,
  Maurin, Naess, Schmitt, Abitbol, Addison, Ade, Alonso, Amiri
  et~al.}}]{aiola:2020}
\bibinfo{author}{\bibfnamefont{S.}~\bibnamefont{Aiola}},
  \bibinfo{author}{\bibfnamefont{E.}~\bibnamefont{Calabrese}},
  \bibinfo{author}{\bibfnamefont{L.}~\bibnamefont{Maurin}},
  \bibinfo{author}{\bibfnamefont{S.}~\bibnamefont{Naess}},
  \bibinfo{author}{\bibfnamefont{B.~L.} \bibnamefont{Schmitt}},
  \bibinfo{author}{\bibfnamefont{M.~H.} \bibnamefont{Abitbol}},
  \bibinfo{author}{\bibfnamefont{G.~E.} \bibnamefont{Addison}},
  \bibinfo{author}{\bibfnamefont{P.~A.} \bibnamefont{Ade}},
  \bibinfo{author}{\bibfnamefont{D.}~\bibnamefont{Alonso}},
  \bibinfo{author}{\bibfnamefont{M.}~\bibnamefont{Amiri}},
  \bibnamefont{et~al.}, \bibinfo{journal}{Journal of Cosmology and
  Astroparticle Physics} \textbf{\bibinfo{volume}{2020}}, \bibinfo{pages}{047}
  (\bibinfo{year}{2020}).

\bibitem[{\citenamefont{{Riess} et~al.}(2011)\citenamefont{{Riess}, {Macri},
  {Casertano}, {Lampeitl}, {Ferguson}, {Filippenko}, {Jha}, {Li}, and
  {Chornock}}}]{Riess:2011}
\bibinfo{author}{\bibfnamefont{A.~G.} \bibnamefont{{Riess}}},
  \bibinfo{author}{\bibfnamefont{L.}~\bibnamefont{{Macri}}},
  \bibinfo{author}{\bibfnamefont{S.}~\bibnamefont{{Casertano}}},
  \bibinfo{author}{\bibfnamefont{H.}~\bibnamefont{{Lampeitl}}},
  \bibinfo{author}{\bibfnamefont{H.~C.} \bibnamefont{{Ferguson}}},
  \bibinfo{author}{\bibfnamefont{A.~V.} \bibnamefont{{Filippenko}}},
  \bibinfo{author}{\bibfnamefont{S.~W.} \bibnamefont{{Jha}}},
  \bibinfo{author}{\bibfnamefont{W.}~\bibnamefont{{Li}}}, \bibnamefont{and}
  \bibinfo{author}{\bibfnamefont{R.}~\bibnamefont{{Chornock}}},
  \bibinfo{journal}{\apj} \textbf{\bibinfo{volume}{730}}, \bibinfo{eid}{119}
  (\bibinfo{year}{2011}), \eprint{1103.2976}.

\bibitem[{\citenamefont{Riess et~al.}(2016)}]{Riess:2016jrr}
\bibinfo{author}{\bibfnamefont{A.~G.} \bibnamefont{Riess}}
  \bibnamefont{et~al.}, \bibinfo{journal}{Astrophys. J.}
  \textbf{\bibinfo{volume}{826}}, \bibinfo{pages}{56} (\bibinfo{year}{2016}),
  \eprint{1604.01424}.

\bibitem[{\citenamefont{Riess et~al.}(2019)\citenamefont{Riess, Casertano,
  Yuan, Macri, and Scolnic}}]{Riess2019}
\bibinfo{author}{\bibfnamefont{A.~G.} \bibnamefont{Riess}},
  \bibinfo{author}{\bibfnamefont{S.}~\bibnamefont{Casertano}},
  \bibinfo{author}{\bibfnamefont{W.}~\bibnamefont{Yuan}},
  \bibinfo{author}{\bibfnamefont{L.~M.} \bibnamefont{Macri}}, \bibnamefont{and}
  \bibinfo{author}{\bibfnamefont{D.}~\bibnamefont{Scolnic}},
  \bibinfo{journal}{Astrophys. J.} \textbf{\bibinfo{volume}{876}},
  \bibinfo{pages}{85} (\bibinfo{year}{2019}), \eprint{1903.07603}.

\bibitem[{\citenamefont{{Riess} et~al.}(2021)\citenamefont{{Riess},
  {Casertano}, {Yuan}, {Bowers}, {Macri}, {Zinn}, and {Scolnic}}}]{riess2021}
\bibinfo{author}{\bibfnamefont{A.~G.} \bibnamefont{{Riess}}},
  \bibinfo{author}{\bibfnamefont{S.}~\bibnamefont{{Casertano}}},
  \bibinfo{author}{\bibfnamefont{W.}~\bibnamefont{{Yuan}}},
  \bibinfo{author}{\bibfnamefont{J.~B.} \bibnamefont{{Bowers}}},
  \bibinfo{author}{\bibfnamefont{L.}~\bibnamefont{{Macri}}},
  \bibinfo{author}{\bibfnamefont{J.~C.} \bibnamefont{{Zinn}}},
  \bibnamefont{and}
  \bibinfo{author}{\bibfnamefont{D.}~\bibnamefont{{Scolnic}}},
  \bibinfo{journal}{\apjl} \textbf{\bibinfo{volume}{908}}, \bibinfo{eid}{L6}
  (\bibinfo{year}{2021}), \eprint{2012.08534}.

\bibitem[{\citenamefont{Riess et~al.}(2021)}]{Riess:2021jrx}
\bibinfo{author}{\bibfnamefont{A.~G.} \bibnamefont{Riess}} \bibnamefont{et~al.}
  (\bibinfo{year}{2021}), \eprint{2112.04510}.

\bibitem[{\citenamefont{Percival et~al.}(2010)\citenamefont{Percival, Reid,
  Eisenstein, Bahcall, Budavari, Frieman, Fukugita, Gunn, Ivezi{\'c}, Knapp
  et~al.}}]{percival:2010baryon}
\bibinfo{author}{\bibfnamefont{W.~J.} \bibnamefont{Percival}},
  \bibinfo{author}{\bibfnamefont{B.~A.} \bibnamefont{Reid}},
  \bibinfo{author}{\bibfnamefont{D.~J.} \bibnamefont{Eisenstein}},
  \bibinfo{author}{\bibfnamefont{N.~A.} \bibnamefont{Bahcall}},
  \bibinfo{author}{\bibfnamefont{T.}~\bibnamefont{Budavari}},
  \bibinfo{author}{\bibfnamefont{J.~A.} \bibnamefont{Frieman}},
  \bibinfo{author}{\bibfnamefont{M.}~\bibnamefont{Fukugita}},
  \bibinfo{author}{\bibfnamefont{J.~E.} \bibnamefont{Gunn}},
  \bibinfo{author}{\bibfnamefont{{\v{Z}}.}~\bibnamefont{Ivezi{\'c}}},
  \bibinfo{author}{\bibfnamefont{G.~R.} \bibnamefont{Knapp}},
  \bibnamefont{et~al.}, \bibinfo{journal}{Monthly Notices of the Royal
  Astronomical Society} \textbf{\bibinfo{volume}{401}}, \bibinfo{pages}{2148}
  (\bibinfo{year}{2010}).

\bibitem[{\citenamefont{Heavens et~al.}(2014)\citenamefont{Heavens, Jimenez,
  and Verde}}]{heavens:2014standard}
\bibinfo{author}{\bibfnamefont{A.}~\bibnamefont{Heavens}},
  \bibinfo{author}{\bibfnamefont{R.}~\bibnamefont{Jimenez}}, \bibnamefont{and}
  \bibinfo{author}{\bibfnamefont{L.}~\bibnamefont{Verde}},
  \bibinfo{journal}{Physical review letters} \textbf{\bibinfo{volume}{113}},
  \bibinfo{pages}{241302} (\bibinfo{year}{2014}).

\bibitem[{\citenamefont{Aubourg et~al.}(2015)\citenamefont{Aubourg, Bailey,
  Bautista, Beutler, Bhardwaj, Bizyaev, Blanton, Blomqvist, Bolton, Bovy
  et~al.}}]{aubourg:2015cosmological}
\bibinfo{author}{\bibfnamefont{{\'E}.}~\bibnamefont{Aubourg}},
  \bibinfo{author}{\bibfnamefont{S.}~\bibnamefont{Bailey}},
  \bibinfo{author}{\bibfnamefont{J.~E.} \bibnamefont{Bautista}},
  \bibinfo{author}{\bibfnamefont{F.}~\bibnamefont{Beutler}},
  \bibinfo{author}{\bibfnamefont{V.}~\bibnamefont{Bhardwaj}},
  \bibinfo{author}{\bibfnamefont{D.}~\bibnamefont{Bizyaev}},
  \bibinfo{author}{\bibfnamefont{M.}~\bibnamefont{Blanton}},
  \bibinfo{author}{\bibfnamefont{M.}~\bibnamefont{Blomqvist}},
  \bibinfo{author}{\bibfnamefont{A.~S.} \bibnamefont{Bolton}},
  \bibinfo{author}{\bibfnamefont{J.}~\bibnamefont{Bovy}}, \bibnamefont{et~al.},
  \bibinfo{journal}{Physical Review D} \textbf{\bibinfo{volume}{92}},
  \bibinfo{pages}{123516} (\bibinfo{year}{2015}).

\bibitem[{\citenamefont{Cuesta et~al.}(2015)\citenamefont{Cuesta, Verde, Riess,
  and Jimenez}}]{cuesta:2015calibrating}
\bibinfo{author}{\bibfnamefont{A.~J.} \bibnamefont{Cuesta}},
  \bibinfo{author}{\bibfnamefont{L.}~\bibnamefont{Verde}},
  \bibinfo{author}{\bibfnamefont{A.}~\bibnamefont{Riess}}, \bibnamefont{and}
  \bibinfo{author}{\bibfnamefont{R.}~\bibnamefont{Jimenez}},
  \bibinfo{journal}{Monthly Notices of the Royal Astronomical Society}
  \textbf{\bibinfo{volume}{448}}, \bibinfo{pages}{3463} (\bibinfo{year}{2015}).

\bibitem[{\citenamefont{Bernal et~al.}(2016)\citenamefont{Bernal, Verde, and
  Riess}}]{Bernal:2016gxb}
\bibinfo{author}{\bibfnamefont{J.~L.} \bibnamefont{Bernal}},
  \bibinfo{author}{\bibfnamefont{L.}~\bibnamefont{Verde}}, \bibnamefont{and}
  \bibinfo{author}{\bibfnamefont{A.~G.} \bibnamefont{Riess}},
  \bibinfo{journal}{JCAP} \textbf{\bibinfo{volume}{1610}}, \bibinfo{pages}{019}
  (\bibinfo{year}{2016}), \eprint{1607.05617}.

\bibitem[{\citenamefont{Verde et~al.}(2017)\citenamefont{Verde, Bernal,
  Heavens, and Jimenez}}]{verde:2017length}
\bibinfo{author}{\bibfnamefont{L.}~\bibnamefont{Verde}},
  \bibinfo{author}{\bibfnamefont{J.~L.} \bibnamefont{Bernal}},
  \bibinfo{author}{\bibfnamefont{A.~F.} \bibnamefont{Heavens}},
  \bibnamefont{and} \bibinfo{author}{\bibfnamefont{R.}~\bibnamefont{Jimenez}},
  \bibinfo{journal}{Monthly Notices of the Royal Astronomical Society}
  \textbf{\bibinfo{volume}{467}}, \bibinfo{pages}{731} (\bibinfo{year}{2017}).

\bibitem[{\citenamefont{{Lemos} et~al.}(2019)\citenamefont{{Lemos}, {Lee},
  {Efstathiou}, and {Gratton}}}]{Lemos:2019}
\bibinfo{author}{\bibfnamefont{P.}~\bibnamefont{{Lemos}}},
  \bibinfo{author}{\bibfnamefont{E.}~\bibnamefont{{Lee}}},
  \bibinfo{author}{\bibfnamefont{G.}~\bibnamefont{{Efstathiou}}},
  \bibnamefont{and}
  \bibinfo{author}{\bibfnamefont{S.}~\bibnamefont{{Gratton}}},
  \bibinfo{journal}{\mnras} \textbf{\bibinfo{volume}{483}},
  \bibinfo{pages}{4803} (\bibinfo{year}{2019}), \eprint{1806.06781}.

\bibitem[{\citenamefont{Abbott et~al.}(2018)}]{DES:2017txv}
\bibinfo{author}{\bibfnamefont{T.~M.~C.} \bibnamefont{Abbott}}
  \bibnamefont{et~al.} (\bibinfo{collaboration}{DES}), \bibinfo{journal}{Mon.
  Not. Roy. Astron. Soc.} \textbf{\bibinfo{volume}{480}}, \bibinfo{pages}{3879}
  (\bibinfo{year}{2018}), \eprint{1711.00403}.

\bibitem[{\citenamefont{D'Amico et~al.}(2020)\citenamefont{D'Amico, Gleyzes,
  Kokron, Markovic, Senatore, Zhang, Beutler, and
  Gil-Mar\'\i{}n}}]{DAmico:2019fhj}
\bibinfo{author}{\bibfnamefont{G.}~\bibnamefont{D'Amico}},
  \bibinfo{author}{\bibfnamefont{J.}~\bibnamefont{Gleyzes}},
  \bibinfo{author}{\bibfnamefont{N.}~\bibnamefont{Kokron}},
  \bibinfo{author}{\bibfnamefont{K.}~\bibnamefont{Markovic}},
  \bibinfo{author}{\bibfnamefont{L.}~\bibnamefont{Senatore}},
  \bibinfo{author}{\bibfnamefont{P.}~\bibnamefont{Zhang}},
  \bibinfo{author}{\bibfnamefont{F.}~\bibnamefont{Beutler}}, \bibnamefont{and}
  \bibinfo{author}{\bibfnamefont{H.}~\bibnamefont{Gil-Mar\'\i{}n}},
  \bibinfo{journal}{JCAP} \textbf{\bibinfo{volume}{05}}, \bibinfo{pages}{005}
  (\bibinfo{year}{2020}), \eprint{1909.05271}.

\bibitem[{\citenamefont{Ivanov et~al.}(2020{\natexlab{a}})\citenamefont{Ivanov,
  Simonovi\'c, and Zaldarriaga}}]{Ivanov:2019pdj}
\bibinfo{author}{\bibfnamefont{M.~M.} \bibnamefont{Ivanov}},
  \bibinfo{author}{\bibfnamefont{M.}~\bibnamefont{Simonovi\'c}},
  \bibnamefont{and}
  \bibinfo{author}{\bibfnamefont{M.}~\bibnamefont{Zaldarriaga}},
  \bibinfo{journal}{JCAP} \textbf{\bibinfo{volume}{05}}, \bibinfo{pages}{042}
  (\bibinfo{year}{2020}{\natexlab{a}}), \eprint{1909.05277}.

\bibitem[{\citenamefont{Colas et~al.}(2020)\citenamefont{Colas, D'amico,
  Senatore, Zhang, and Beutler}}]{Colas:2019ret}
\bibinfo{author}{\bibfnamefont{T.}~\bibnamefont{Colas}},
  \bibinfo{author}{\bibfnamefont{G.}~\bibnamefont{D'amico}},
  \bibinfo{author}{\bibfnamefont{L.}~\bibnamefont{Senatore}},
  \bibinfo{author}{\bibfnamefont{P.}~\bibnamefont{Zhang}}, \bibnamefont{and}
  \bibinfo{author}{\bibfnamefont{F.}~\bibnamefont{Beutler}},
  \bibinfo{journal}{JCAP} \textbf{\bibinfo{volume}{06}}, \bibinfo{pages}{001}
  (\bibinfo{year}{2020}), \eprint{1909.07951}.

\bibitem[{\citenamefont{Philcox et~al.}(2020)\citenamefont{Philcox, Ivanov,
  Simonovi\'c, and Zaldarriaga}}]{Philcox:2020vvt}
\bibinfo{author}{\bibfnamefont{O.~H.~E.} \bibnamefont{Philcox}},
  \bibinfo{author}{\bibfnamefont{M.~M.} \bibnamefont{Ivanov}},
  \bibinfo{author}{\bibfnamefont{M.}~\bibnamefont{Simonovi\'c}},
  \bibnamefont{and}
  \bibinfo{author}{\bibfnamefont{M.}~\bibnamefont{Zaldarriaga}},
  \bibinfo{journal}{JCAP} \textbf{\bibinfo{volume}{05}}, \bibinfo{pages}{032}
  (\bibinfo{year}{2020}), \eprint{2002.04035}.

\bibitem[{\citenamefont{Zhang et~al.}(2022)\citenamefont{Zhang, D'Amico,
  Senatore, Zhao, and Cai}}]{Zhang:2021yna}
\bibinfo{author}{\bibfnamefont{P.}~\bibnamefont{Zhang}},
  \bibinfo{author}{\bibfnamefont{G.}~\bibnamefont{D'Amico}},
  \bibinfo{author}{\bibfnamefont{L.}~\bibnamefont{Senatore}},
  \bibinfo{author}{\bibfnamefont{C.}~\bibnamefont{Zhao}}, \bibnamefont{and}
  \bibinfo{author}{\bibfnamefont{Y.}~\bibnamefont{Cai}},
  \bibinfo{journal}{JCAP} \textbf{\bibinfo{volume}{02}}, \bibinfo{pages}{036}
  (\bibinfo{year}{2022}), \eprint{2110.07539}.

\bibitem[{\citenamefont{{Freedman} et~al.}(2012)\citenamefont{{Freedman},
  {Madore}, {Scowcroft}, {Burns}, {Monson}, {Persson}, {Seibert}, and
  {Rigby}}}]{freedman2012}
\bibinfo{author}{\bibfnamefont{W.~L.} \bibnamefont{{Freedman}}},
  \bibinfo{author}{\bibfnamefont{B.~F.} \bibnamefont{{Madore}}},
  \bibinfo{author}{\bibfnamefont{V.}~\bibnamefont{{Scowcroft}}},
  \bibinfo{author}{\bibfnamefont{C.}~\bibnamefont{{Burns}}},
  \bibinfo{author}{\bibfnamefont{A.}~\bibnamefont{{Monson}}},
  \bibinfo{author}{\bibfnamefont{S.~E.} \bibnamefont{{Persson}}},
  \bibinfo{author}{\bibfnamefont{M.}~\bibnamefont{{Seibert}}},
  \bibnamefont{and} \bibinfo{author}{\bibfnamefont{J.}~\bibnamefont{{Rigby}}},
  \bibinfo{journal}{\apj} \textbf{\bibinfo{volume}{758}}, \bibinfo{eid}{24}
  (\bibinfo{year}{2012}), \eprint{1208.3281}.

\bibitem[{\citenamefont{Suyu et~al.}(2017)}]{Suyu:2016qxx}
\bibinfo{author}{\bibfnamefont{S.~H.} \bibnamefont{Suyu}} \bibnamefont{et~al.},
  \bibinfo{journal}{Mon. Not. Roy. Astron. Soc.}
  \textbf{\bibinfo{volume}{468}}, \bibinfo{pages}{2590} (\bibinfo{year}{2017}),
  \eprint{1607.00017}.

\bibitem[{\citenamefont{{Birrer} et~al.}(2019)\citenamefont{{Birrer}, {Treu},
  {Rusu}, {Bonvin}, {Fassnacht}, {Chan}, {Agnello}, {Shajib}, {Chen}, {Auger}
  et~al.}}]{Birrer:2018vtm}
\bibinfo{author}{\bibfnamefont{S.}~\bibnamefont{{Birrer}}},
  \bibinfo{author}{\bibfnamefont{T.}~\bibnamefont{{Treu}}},
  \bibinfo{author}{\bibfnamefont{C.~E.} \bibnamefont{{Rusu}}},
  \bibinfo{author}{\bibfnamefont{V.}~\bibnamefont{{Bonvin}}},
  \bibinfo{author}{\bibfnamefont{C.~D.} \bibnamefont{{Fassnacht}}},
  \bibinfo{author}{\bibfnamefont{J.~H.~H.} \bibnamefont{{Chan}}},
  \bibinfo{author}{\bibfnamefont{A.}~\bibnamefont{{Agnello}}},
  \bibinfo{author}{\bibfnamefont{A.~J.} \bibnamefont{{Shajib}}},
  \bibinfo{author}{\bibfnamefont{G.~C.~F.} \bibnamefont{{Chen}}},
  \bibinfo{author}{\bibfnamefont{M.}~\bibnamefont{{Auger}}},
  \bibnamefont{et~al.}, \bibinfo{journal}{\mnras}
  \textbf{\bibinfo{volume}{484}}, \bibinfo{pages}{4726} (\bibinfo{year}{2019}),
  \eprint{1809.01274}.

\bibitem[{\citenamefont{Wong et~al.}(2020)}]{Wong:2019kwg}
\bibinfo{author}{\bibfnamefont{K.~C.} \bibnamefont{Wong}} \bibnamefont{et~al.},
  \bibinfo{journal}{Mon. Not. Roy. Astron. Soc.}
  \textbf{\bibinfo{volume}{498}}, \bibinfo{pages}{1420} (\bibinfo{year}{2020}),
  \eprint{1907.04869}.

\bibitem[{\citenamefont{Huang et~al.}(2019)\citenamefont{Huang, Riess, Yuan,
  Macri, Zakamska, Casertano, Whitelock, Hoffmann, Filippenko, and
  Scolnic}}]{Huang:2019yhh}
\bibinfo{author}{\bibfnamefont{C.~D.} \bibnamefont{Huang}},
  \bibinfo{author}{\bibfnamefont{A.~G.} \bibnamefont{Riess}},
  \bibinfo{author}{\bibfnamefont{W.}~\bibnamefont{Yuan}},
  \bibinfo{author}{\bibfnamefont{L.~M.} \bibnamefont{Macri}},
  \bibinfo{author}{\bibfnamefont{N.~L.} \bibnamefont{Zakamska}},
  \bibinfo{author}{\bibfnamefont{S.}~\bibnamefont{Casertano}},
  \bibinfo{author}{\bibfnamefont{P.~A.} \bibnamefont{Whitelock}},
  \bibinfo{author}{\bibfnamefont{S.~L.} \bibnamefont{Hoffmann}},
  \bibinfo{author}{\bibfnamefont{A.~V.} \bibnamefont{Filippenko}},
  \bibnamefont{and} \bibinfo{author}{\bibfnamefont{D.}~\bibnamefont{Scolnic}}
  (\bibinfo{year}{2019}), \eprint{1908.10883}.

\bibitem[{\citenamefont{Kourkchi et~al.}(2020)\citenamefont{Kourkchi, Tully,
  Anand, Courtois, Dupuy, Neill, Rizzi, and Seibert}}]{Kourkchi:2020iyz}
\bibinfo{author}{\bibfnamefont{E.}~\bibnamefont{Kourkchi}},
  \bibinfo{author}{\bibfnamefont{R.~B.} \bibnamefont{Tully}},
  \bibinfo{author}{\bibfnamefont{G.~S.} \bibnamefont{Anand}},
  \bibinfo{author}{\bibfnamefont{H.~M.} \bibnamefont{Courtois}},
  \bibinfo{author}{\bibfnamefont{A.}~\bibnamefont{Dupuy}},
  \bibinfo{author}{\bibfnamefont{J.~D.} \bibnamefont{Neill}},
  \bibinfo{author}{\bibfnamefont{L.}~\bibnamefont{Rizzi}}, \bibnamefont{and}
  \bibinfo{author}{\bibfnamefont{M.}~\bibnamefont{Seibert}},
  \bibinfo{journal}{Astrophys. J.} \textbf{\bibinfo{volume}{896}},
  \bibinfo{pages}{3} (\bibinfo{year}{2020}), \eprint{2004.14499}.

\bibitem[{\citenamefont{Reid et~al.}(2019)\citenamefont{Reid, Pesce, and
  Riess}}]{Reid:2019tiq}
\bibinfo{author}{\bibfnamefont{M.~J.} \bibnamefont{Reid}},
  \bibinfo{author}{\bibfnamefont{D.~W.} \bibnamefont{Pesce}}, \bibnamefont{and}
  \bibinfo{author}{\bibfnamefont{A.~G.} \bibnamefont{Riess}},
  \bibinfo{journal}{Astrophys. J. Lett.} \textbf{\bibinfo{volume}{886}},
  \bibinfo{pages}{L27} (\bibinfo{year}{2019}), \eprint{1908.05625}.

\bibitem[{\citenamefont{Freedman et~al.}(2020)\citenamefont{Freedman, Madore,
  Hoyt, Jang, Beaton, Lee, Monson, Neeley, and Rich}}]{Freedman:2020dne}
\bibinfo{author}{\bibfnamefont{W.~L.} \bibnamefont{Freedman}},
  \bibinfo{author}{\bibfnamefont{B.~F.} \bibnamefont{Madore}},
  \bibinfo{author}{\bibfnamefont{T.}~\bibnamefont{Hoyt}},
  \bibinfo{author}{\bibfnamefont{I.~S.} \bibnamefont{Jang}},
  \bibinfo{author}{\bibfnamefont{R.}~\bibnamefont{Beaton}},
  \bibinfo{author}{\bibfnamefont{M.~G.} \bibnamefont{Lee}},
  \bibinfo{author}{\bibfnamefont{A.}~\bibnamefont{Monson}},
  \bibinfo{author}{\bibfnamefont{J.}~\bibnamefont{Neeley}}, \bibnamefont{and}
  \bibinfo{author}{\bibfnamefont{J.}~\bibnamefont{Rich}}
  (\bibinfo{year}{2020}), \eprint{2002.01550}.

\bibitem[{\citenamefont{Freedman}(2021)}]{Freedman:2021ahq}
\bibinfo{author}{\bibfnamefont{W.~L.} \bibnamefont{Freedman}}
  (\bibinfo{year}{2021}), \eprint{2106.15656}.

\bibitem[{\citenamefont{Pesce et~al.}(2020)}]{Pesce:2020xfe}
\bibinfo{author}{\bibfnamefont{D.~W.} \bibnamefont{Pesce}}
  \bibnamefont{et~al.}, \bibinfo{journal}{Astrophys. J. Lett.}
  \textbf{\bibinfo{volume}{891}}, \bibinfo{pages}{L1} (\bibinfo{year}{2020}),
  \eprint{2001.09213}.

\bibitem[{\citenamefont{Khetan et~al.}(2021)}]{Khetan:2020hmh}
\bibinfo{author}{\bibfnamefont{N.}~\bibnamefont{Khetan}} \bibnamefont{et~al.},
  \bibinfo{journal}{Astron. Astrophys.} \textbf{\bibinfo{volume}{647}},
  \bibinfo{pages}{A72} (\bibinfo{year}{2021}), \eprint{2008.07754}.

\bibitem[{\citenamefont{Blakeslee et~al.}(2021)\citenamefont{Blakeslee, Jensen,
  Ma, Milne, and Greene}}]{Blakeslee:2021rqi}
\bibinfo{author}{\bibfnamefont{J.~P.} \bibnamefont{Blakeslee}},
  \bibinfo{author}{\bibfnamefont{J.~B.} \bibnamefont{Jensen}},
  \bibinfo{author}{\bibfnamefont{C.-P.} \bibnamefont{Ma}},
  \bibinfo{author}{\bibfnamefont{P.~A.} \bibnamefont{Milne}}, \bibnamefont{and}
  \bibinfo{author}{\bibfnamefont{J.~E.} \bibnamefont{Greene}},
  \bibinfo{journal}{Astrophys. J.} \textbf{\bibinfo{volume}{911}},
  \bibinfo{pages}{65} (\bibinfo{year}{2021}), \eprint{2101.02221}.

\bibitem[{\citenamefont{Birrer et~al.}(2020)}]{Birrer:2020tax}
\bibinfo{author}{\bibfnamefont{S.}~\bibnamefont{Birrer}} \bibnamefont{et~al.},
  \bibinfo{journal}{Astron. Astrophys.} \textbf{\bibinfo{volume}{643}},
  \bibinfo{pages}{A165} (\bibinfo{year}{2020}), \eprint{2007.02941}.

\bibitem[{\citenamefont{Di~Valentino}(2021)}]{DiValentino:2020vnx}
\bibinfo{author}{\bibfnamefont{E.}~\bibnamefont{Di~Valentino}},
  \bibinfo{journal}{Mon. Not. Roy. Astron. Soc.}
  \textbf{\bibinfo{volume}{502}}, \bibinfo{pages}{2065} (\bibinfo{year}{2021}),
  \eprint{2011.00246}.

\bibitem[{\citenamefont{Beaton et~al.}(2016)}]{Beaton:2016nsw}
\bibinfo{author}{\bibfnamefont{R.~L.} \bibnamefont{Beaton}}
  \bibnamefont{et~al.}, \bibinfo{journal}{Astrophys. J.}
  \textbf{\bibinfo{volume}{832}}, \bibinfo{pages}{210} (\bibinfo{year}{2016}),
  \eprint{1604.01788}.

\bibitem[{\citenamefont{Hatt et~al.}(2017)}]{Hatt:2017rxl}
\bibinfo{author}{\bibfnamefont{D.}~\bibnamefont{Hatt}} \bibnamefont{et~al.},
  \bibinfo{journal}{Astrophys. J.} \textbf{\bibinfo{volume}{845}},
  \bibinfo{pages}{146} (\bibinfo{year}{2017}), \eprint{1703.06468}.

\bibitem[{\citenamefont{Hatt et~al.}(2018{\natexlab{a}})}]{Hatt:2018opj}
\bibinfo{author}{\bibfnamefont{D.}~\bibnamefont{Hatt}} \bibnamefont{et~al.},
  \bibinfo{journal}{Astrophys. J.} \textbf{\bibinfo{volume}{861}},
  \bibinfo{pages}{104} (\bibinfo{year}{2018}{\natexlab{a}}),
  \eprint{1806.02900}.

\bibitem[{\citenamefont{Hatt et~al.}(2018{\natexlab{b}})}]{Hatt:2018zfv}
\bibinfo{author}{\bibfnamefont{D.}~\bibnamefont{Hatt}} \bibnamefont{et~al.},
  \bibinfo{journal}{Astrophys. J.} \textbf{\bibinfo{volume}{866}},
  \bibinfo{pages}{145} (\bibinfo{year}{2018}{\natexlab{b}}),
  \eprint{1809.01741}.

\bibitem[{\citenamefont{Burns et~al.}(2018)}]{CSP:2018rag}
\bibinfo{author}{\bibfnamefont{C.~R.} \bibnamefont{Burns}} \bibnamefont{et~al.}
  (\bibinfo{collaboration}{CSP}), \bibinfo{journal}{Astrophys. J.}
  \textbf{\bibinfo{volume}{869}}, \bibinfo{pages}{56} (\bibinfo{year}{2018}),
  \eprint{1809.06381}.

\bibitem[{\citenamefont{Hoyt et~al.}(2019)\citenamefont{Hoyt, Freedman, Madore,
  Hatt, Beaton, Jang, Lee, Monson, Neeley, Rich et~al.}}]{Hoyt_2019}
\bibinfo{author}{\bibfnamefont{T.~J.} \bibnamefont{Hoyt}},
  \bibinfo{author}{\bibfnamefont{W.~L.} \bibnamefont{Freedman}},
  \bibinfo{author}{\bibfnamefont{B.~F.} \bibnamefont{Madore}},
  \bibinfo{author}{\bibfnamefont{D.}~\bibnamefont{Hatt}},
  \bibinfo{author}{\bibfnamefont{R.~L.} \bibnamefont{Beaton}},
  \bibinfo{author}{\bibfnamefont{I.~S.} \bibnamefont{Jang}},
  \bibinfo{author}{\bibfnamefont{M.~G.} \bibnamefont{Lee}},
  \bibinfo{author}{\bibfnamefont{A.~J.} \bibnamefont{Monson}},
  \bibinfo{author}{\bibfnamefont{J.~R.} \bibnamefont{Neeley}},
  \bibinfo{author}{\bibfnamefont{J.~A.} \bibnamefont{Rich}},
  \bibnamefont{et~al.}, \bibinfo{journal}{The Astrophysical Journal}
  \textbf{\bibinfo{volume}{882}}, \bibinfo{pages}{150} (\bibinfo{year}{2019}),
  ISSN \bibinfo{issn}{1538-4357}.

\bibitem[{\citenamefont{Beaton et~al.}(2019)\citenamefont{Beaton, Seibert,
  Hatt, Freedman, Hoyt, Jang, Lee, Madore, Monson, Neeley
  et~al.}}]{Beaton_2019}
\bibinfo{author}{\bibfnamefont{R.~L.} \bibnamefont{Beaton}},
  \bibinfo{author}{\bibfnamefont{M.}~\bibnamefont{Seibert}},
  \bibinfo{author}{\bibfnamefont{D.}~\bibnamefont{Hatt}},
  \bibinfo{author}{\bibfnamefont{W.~L.} \bibnamefont{Freedman}},
  \bibinfo{author}{\bibfnamefont{T.~J.} \bibnamefont{Hoyt}},
  \bibinfo{author}{\bibfnamefont{I.~S.} \bibnamefont{Jang}},
  \bibinfo{author}{\bibfnamefont{M.~G.} \bibnamefont{Lee}},
  \bibinfo{author}{\bibfnamefont{B.~F.} \bibnamefont{Madore}},
  \bibinfo{author}{\bibfnamefont{A.~J.} \bibnamefont{Monson}},
  \bibinfo{author}{\bibfnamefont{J.~R.} \bibnamefont{Neeley}},
  \bibnamefont{et~al.}, \bibinfo{journal}{The Astrophysical Journal}
  \textbf{\bibinfo{volume}{885}}, \bibinfo{pages}{141} (\bibinfo{year}{2019}),
  ISSN \bibinfo{issn}{1538-4357}.

\bibitem[{\citenamefont{{Freedman} et~al.}(2019)\citenamefont{{Freedman},
  {Madore}, {Hatt}, {Hoyt}, {Jang}, {Beaton}, {Burns}, {Lee}, {Monson},
  {Neeley} et~al.}}]{freedman2019}
\bibinfo{author}{\bibfnamefont{W.~L.} \bibnamefont{{Freedman}}},
  \bibinfo{author}{\bibfnamefont{B.~F.} \bibnamefont{{Madore}}},
  \bibinfo{author}{\bibfnamefont{D.}~\bibnamefont{{Hatt}}},
  \bibinfo{author}{\bibfnamefont{T.~J.} \bibnamefont{{Hoyt}}},
  \bibinfo{author}{\bibfnamefont{I.~S.} \bibnamefont{{Jang}}},
  \bibinfo{author}{\bibfnamefont{R.~L.} \bibnamefont{{Beaton}}},
  \bibinfo{author}{\bibfnamefont{C.~R.} \bibnamefont{{Burns}}},
  \bibinfo{author}{\bibfnamefont{M.~G.} \bibnamefont{{Lee}}},
  \bibinfo{author}{\bibfnamefont{A.~J.} \bibnamefont{{Monson}}},
  \bibinfo{author}{\bibfnamefont{J.~R.} \bibnamefont{{Neeley}}},
  \bibnamefont{et~al.}, \bibinfo{journal}{\apj} \textbf{\bibinfo{volume}{882}},
  \bibinfo{eid}{34} (\bibinfo{year}{2019}), \eprint{1907.05922}.

\bibitem[{\citenamefont{Jang et~al.}(2021)\citenamefont{Jang, Hoyt, Beaton,
  Freedman, Madore, Lee, Neeley, Monson, Rich, and Seibert}}]{Jang_2021}
\bibinfo{author}{\bibfnamefont{I.~S.} \bibnamefont{Jang}},
  \bibinfo{author}{\bibfnamefont{T.~J.} \bibnamefont{Hoyt}},
  \bibinfo{author}{\bibfnamefont{R.~L.} \bibnamefont{Beaton}},
  \bibinfo{author}{\bibfnamefont{W.~L.} \bibnamefont{Freedman}},
  \bibinfo{author}{\bibfnamefont{B.~F.} \bibnamefont{Madore}},
  \bibinfo{author}{\bibfnamefont{M.~G.} \bibnamefont{Lee}},
  \bibinfo{author}{\bibfnamefont{J.~R.} \bibnamefont{Neeley}},
  \bibinfo{author}{\bibfnamefont{A.~J.} \bibnamefont{Monson}},
  \bibinfo{author}{\bibfnamefont{J.~A.} \bibnamefont{Rich}}, \bibnamefont{and}
  \bibinfo{author}{\bibfnamefont{M.}~\bibnamefont{Seibert}},
  \bibinfo{journal}{The Astrophysical Journal} \textbf{\bibinfo{volume}{906}},
  \bibinfo{pages}{125} (\bibinfo{year}{2021}), ISSN \bibinfo{issn}{1538-4357}.

\bibitem[{\citenamefont{Hoyt et~al.}(2021)\citenamefont{Hoyt, Beaton, Freedman,
  Jang, Lee, Madore, Monson, Neeley, Rich, and Seibert}}]{Hoyt_2021}
\bibinfo{author}{\bibfnamefont{T.~J.} \bibnamefont{Hoyt}},
  \bibinfo{author}{\bibfnamefont{R.~L.} \bibnamefont{Beaton}},
  \bibinfo{author}{\bibfnamefont{W.~L.} \bibnamefont{Freedman}},
  \bibinfo{author}{\bibfnamefont{I.~S.} \bibnamefont{Jang}},
  \bibinfo{author}{\bibfnamefont{M.~G.} \bibnamefont{Lee}},
  \bibinfo{author}{\bibfnamefont{B.~F.} \bibnamefont{Madore}},
  \bibinfo{author}{\bibfnamefont{A.~J.} \bibnamefont{Monson}},
  \bibinfo{author}{\bibfnamefont{J.~R.} \bibnamefont{Neeley}},
  \bibinfo{author}{\bibfnamefont{J.~A.} \bibnamefont{Rich}}, \bibnamefont{and}
  \bibinfo{author}{\bibfnamefont{M.}~\bibnamefont{Seibert}},
  \bibinfo{journal}{The Astrophysical Journal} \textbf{\bibinfo{volume}{915}},
  \bibinfo{pages}{34} (\bibinfo{year}{2021}), ISSN \bibinfo{issn}{1538-4357}.

\bibitem[{\citenamefont{{Shah} et~al.}(2021)\citenamefont{{Shah}, {Lemos}, and
  {Lahav}}}]{Paul:2021abc}
\bibinfo{author}{\bibfnamefont{P.}~\bibnamefont{{Shah}}},
  \bibinfo{author}{\bibfnamefont{P.}~\bibnamefont{{Lemos}}}, \bibnamefont{and}
  \bibinfo{author}{\bibfnamefont{O.}~\bibnamefont{{Lahav}}},
  \bibinfo{journal}{arXiv e-prints} \bibinfo{eid}{arXiv:2109.01161}
  (\bibinfo{year}{2021}), \eprint{2109.01161}.

\bibitem[{\citenamefont{Knox and Millea}(2020)}]{knox20}
\bibinfo{author}{\bibfnamefont{L.}~\bibnamefont{Knox}} \bibnamefont{and}
  \bibinfo{author}{\bibfnamefont{M.}~\bibnamefont{Millea}},
  \bibinfo{journal}{Phys. Rev. D} \textbf{\bibinfo{volume}{101}},
  \bibinfo{pages}{043533} (\bibinfo{year}{2020}).

\bibitem[{\citenamefont{Di~Valentino et~al.}(2021)\citenamefont{Di~Valentino,
  Mena, Pan, Visinelli, Yang, Melchiorri, Mota, Riess, and
  Silk}}]{divalentino21}
\bibinfo{author}{\bibfnamefont{E.}~\bibnamefont{Di~Valentino}},
  \bibinfo{author}{\bibfnamefont{O.}~\bibnamefont{Mena}},
  \bibinfo{author}{\bibfnamefont{S.}~\bibnamefont{Pan}},
  \bibinfo{author}{\bibfnamefont{L.}~\bibnamefont{Visinelli}},
  \bibinfo{author}{\bibfnamefont{W.}~\bibnamefont{Yang}},
  \bibinfo{author}{\bibfnamefont{A.}~\bibnamefont{Melchiorri}},
  \bibinfo{author}{\bibfnamefont{D.~F.} \bibnamefont{Mota}},
  \bibinfo{author}{\bibfnamefont{A.~G.} \bibnamefont{Riess}}, \bibnamefont{and}
  \bibinfo{author}{\bibfnamefont{J.}~\bibnamefont{Silk}}
  (\bibinfo{year}{2021}), \eprint{2103.01183}.

\bibitem[{\citenamefont{Sch\"oneberg et~al.}(2021)\citenamefont{Sch\"oneberg,
  Abell\'an, S\'anchez, Witte, Poulin, and Lesgourgues}}]{Schoneberg:2021qvd}
\bibinfo{author}{\bibfnamefont{N.}~\bibnamefont{Sch\"oneberg}},
  \bibinfo{author}{\bibfnamefont{G.~F.} \bibnamefont{Abell\'an}},
  \bibinfo{author}{\bibfnamefont{A.~P.} \bibnamefont{S\'anchez}},
  \bibinfo{author}{\bibfnamefont{S.~J.} \bibnamefont{Witte}},
  \bibinfo{author}{\bibfnamefont{c.~V.} \bibnamefont{Poulin}},
  \bibnamefont{and}
  \bibinfo{author}{\bibfnamefont{J.}~\bibnamefont{Lesgourgues}}
  (\bibinfo{year}{2021}), \eprint{2107.10291}.

\bibitem[{Note1()}]{Note1}
Note1, \bibinfo{note}{redshift is often used as a timelike variable in
  cosmology since the light that arrives here from a given event on our past
  light cone is stretched by the expansion by a factor of $1+z$, providing a
  monotonic relationship between time of emission and redshift $z$ in a
  continuously expanding universe.}

\bibitem[{\citenamefont{Zahn and Zaldarriaga}(2003)}]{Zahn_2003}
\bibinfo{author}{\bibfnamefont{O.}~\bibnamefont{Zahn}} \bibnamefont{and}
  \bibinfo{author}{\bibfnamefont{M.}~\bibnamefont{Zaldarriaga}},
  \bibinfo{journal}{Physical Review D} \textbf{\bibinfo{volume}{67}}
  (\bibinfo{year}{2003}), ISSN \bibinfo{issn}{1089-4918}.

\bibitem[{Note2()}]{Note2}
Note2, \bibinfo{note}{models have also been proposed that contain multiple
  copies of the SM, see e.g.~Refs.~\cite
  {Dvali:2007hz,Dvali:2009ne,Arkani-Hamed:2016rle}.}

\bibitem[{\citenamefont{Chacko et~al.}(2006{\natexlab{a}})\citenamefont{Chacko,
  Goh, and Harnik}}]{Chacko:2005pe}
\bibinfo{author}{\bibfnamefont{Z.}~\bibnamefont{Chacko}},
  \bibinfo{author}{\bibfnamefont{H.-S.} \bibnamefont{Goh}}, \bibnamefont{and}
  \bibinfo{author}{\bibfnamefont{R.}~\bibnamefont{Harnik}},
  \bibinfo{journal}{Phys. Rev. Lett.} \textbf{\bibinfo{volume}{96}},
  \bibinfo{pages}{231802} (\bibinfo{year}{2006}{\natexlab{a}}),
  \eprint{hep-ph/0506256}.

\bibitem[{\citenamefont{Chacko et~al.}(2006{\natexlab{b}})\citenamefont{Chacko,
  Nomura, Papucci, and Perez}}]{Chacko:2005vw}
\bibinfo{author}{\bibfnamefont{Z.}~\bibnamefont{Chacko}},
  \bibinfo{author}{\bibfnamefont{Y.}~\bibnamefont{Nomura}},
  \bibinfo{author}{\bibfnamefont{M.}~\bibnamefont{Papucci}}, \bibnamefont{and}
  \bibinfo{author}{\bibfnamefont{G.}~\bibnamefont{Perez}},
  \bibinfo{journal}{JHEP} \textbf{\bibinfo{volume}{01}}, \bibinfo{pages}{126}
  (\bibinfo{year}{2006}{\natexlab{b}}), \eprint{hep-ph/0510273}.

\bibitem[{\citenamefont{Chacko et~al.}(2006{\natexlab{c}})\citenamefont{Chacko,
  Goh, and Harnik}}]{Chacko:2005un}
\bibinfo{author}{\bibfnamefont{Z.}~\bibnamefont{Chacko}},
  \bibinfo{author}{\bibfnamefont{H.-S.} \bibnamefont{Goh}}, \bibnamefont{and}
  \bibinfo{author}{\bibfnamefont{R.}~\bibnamefont{Harnik}},
  \bibinfo{journal}{JHEP} \textbf{\bibinfo{volume}{01}}, \bibinfo{pages}{108}
  (\bibinfo{year}{2006}{\natexlab{c}}), \eprint{hep-ph/0512088}.

\bibitem[{\citenamefont{Barbieri et~al.}(2005)\citenamefont{Barbieri, Gregoire,
  and Hall}}]{Barbieri:2005ri}
\bibinfo{author}{\bibfnamefont{R.}~\bibnamefont{Barbieri}},
  \bibinfo{author}{\bibfnamefont{T.}~\bibnamefont{Gregoire}}, \bibnamefont{and}
  \bibinfo{author}{\bibfnamefont{L.~J.} \bibnamefont{Hall}}
  (\bibinfo{year}{2005}), \eprint{hep-ph/0509242}.

\bibitem[{\citenamefont{Craig and Howe}(2014)}]{Craig:2013fga}
\bibinfo{author}{\bibfnamefont{N.}~\bibnamefont{Craig}} \bibnamefont{and}
  \bibinfo{author}{\bibfnamefont{K.}~\bibnamefont{Howe}},
  \bibinfo{journal}{JHEP} \textbf{\bibinfo{volume}{03}}, \bibinfo{pages}{140}
  (\bibinfo{year}{2014}), \eprint{1312.1341}.

\bibitem[{\citenamefont{Craig et~al.}(2015)\citenamefont{Craig, Katz,
  Strassler, and Sundrum}}]{Craig:2015pha}
\bibinfo{author}{\bibfnamefont{N.}~\bibnamefont{Craig}},
  \bibinfo{author}{\bibfnamefont{A.}~\bibnamefont{Katz}},
  \bibinfo{author}{\bibfnamefont{M.}~\bibnamefont{Strassler}},
  \bibnamefont{and} \bibinfo{author}{\bibfnamefont{R.}~\bibnamefont{Sundrum}},
  \bibinfo{journal}{JHEP} \textbf{\bibinfo{volume}{07}}, \bibinfo{pages}{105}
  (\bibinfo{year}{2015}), \eprint{1501.05310}.

\bibitem[{\citenamefont{Garcia~Garcia et~al.}(2015)\citenamefont{Garcia~Garcia,
  Lasenby, and March-Russell}}]{GarciaGarcia:2015fol}
\bibinfo{author}{\bibfnamefont{I.}~\bibnamefont{Garcia~Garcia}},
  \bibinfo{author}{\bibfnamefont{R.}~\bibnamefont{Lasenby}}, \bibnamefont{and}
  \bibinfo{author}{\bibfnamefont{J.}~\bibnamefont{March-Russell}},
  \bibinfo{journal}{Phys. Rev. D} \textbf{\bibinfo{volume}{92}},
  \bibinfo{pages}{055034} (\bibinfo{year}{2015}), \eprint{1505.07109}.

\bibitem[{\citenamefont{Craig and Katz}(2015)}]{Craig:2015xla}
\bibinfo{author}{\bibfnamefont{N.}~\bibnamefont{Craig}} \bibnamefont{and}
  \bibinfo{author}{\bibfnamefont{A.}~\bibnamefont{Katz}},
  \bibinfo{journal}{JCAP} \textbf{\bibinfo{volume}{10}}, \bibinfo{pages}{054}
  (\bibinfo{year}{2015}), \eprint{1505.07113}.

\bibitem[{\citenamefont{Farina}(2015)}]{Farina:2015uea}
\bibinfo{author}{\bibfnamefont{M.}~\bibnamefont{Farina}},
  \bibinfo{journal}{JCAP} \textbf{\bibinfo{volume}{11}}, \bibinfo{pages}{017}
  (\bibinfo{year}{2015}), \eprint{1506.03520}.

\bibitem[{\citenamefont{Farina et~al.}(2016)\citenamefont{Farina, Monteux, and
  Shin}}]{Farina:2016ndq}
\bibinfo{author}{\bibfnamefont{M.}~\bibnamefont{Farina}},
  \bibinfo{author}{\bibfnamefont{A.}~\bibnamefont{Monteux}}, \bibnamefont{and}
  \bibinfo{author}{\bibfnamefont{C.~S.} \bibnamefont{Shin}},
  \bibinfo{journal}{Phys. Rev. D} \textbf{\bibinfo{volume}{94}},
  \bibinfo{pages}{035017} (\bibinfo{year}{2016}), \eprint{1604.08211}.

\bibitem[{\citenamefont{Prilepina and Tsai}(2017)}]{Prilepina:2016rlq}
\bibinfo{author}{\bibfnamefont{V.}~\bibnamefont{Prilepina}} \bibnamefont{and}
  \bibinfo{author}{\bibfnamefont{Y.}~\bibnamefont{Tsai}},
  \bibinfo{journal}{JHEP} \textbf{\bibinfo{volume}{09}}, \bibinfo{pages}{033}
  (\bibinfo{year}{2017}), \eprint{1611.05879}.

\bibitem[{\citenamefont{Barbieri et~al.}(2016)\citenamefont{Barbieri, Hall, and
  Harigaya}}]{Barbieri:2016zxn}
\bibinfo{author}{\bibfnamefont{R.}~\bibnamefont{Barbieri}},
  \bibinfo{author}{\bibfnamefont{L.~J.} \bibnamefont{Hall}}, \bibnamefont{and}
  \bibinfo{author}{\bibfnamefont{K.}~\bibnamefont{Harigaya}},
  \bibinfo{journal}{JHEP} \textbf{\bibinfo{volume}{11}}, \bibinfo{pages}{172}
  (\bibinfo{year}{2016}), \eprint{1609.05589}.

\bibitem[{\citenamefont{Craig et~al.}(2017)\citenamefont{Craig, Koren, and
  Trott}}]{Craig:2016lyx}
\bibinfo{author}{\bibfnamefont{N.}~\bibnamefont{Craig}},
  \bibinfo{author}{\bibfnamefont{S.}~\bibnamefont{Koren}}, \bibnamefont{and}
  \bibinfo{author}{\bibfnamefont{T.}~\bibnamefont{Trott}},
  \bibinfo{journal}{JHEP} \textbf{\bibinfo{volume}{05}}, \bibinfo{pages}{038}
  (\bibinfo{year}{2017}), \eprint{1611.07977}.

\bibitem[{\citenamefont{Berger et~al.}(2016)\citenamefont{Berger, Jedamzik, and
  Walker}}]{Berger:2016vxi}
\bibinfo{author}{\bibfnamefont{J.}~\bibnamefont{Berger}},
  \bibinfo{author}{\bibfnamefont{K.}~\bibnamefont{Jedamzik}}, \bibnamefont{and}
  \bibinfo{author}{\bibfnamefont{D.~G.~E.} \bibnamefont{Walker}},
  \bibinfo{journal}{JCAP} \textbf{\bibinfo{volume}{11}}, \bibinfo{pages}{032}
  (\bibinfo{year}{2016}), \eprint{1605.07195}.

\bibitem[{\citenamefont{Chacko et~al.}(2017)\citenamefont{Chacko, Craig, Fox,
  and Harnik}}]{Chacko:2016hvu}
\bibinfo{author}{\bibfnamefont{Z.}~\bibnamefont{Chacko}},
  \bibinfo{author}{\bibfnamefont{N.}~\bibnamefont{Craig}},
  \bibinfo{author}{\bibfnamefont{P.~J.} \bibnamefont{Fox}}, \bibnamefont{and}
  \bibinfo{author}{\bibfnamefont{R.}~\bibnamefont{Harnik}},
  \bibinfo{journal}{JHEP} \textbf{\bibinfo{volume}{07}}, \bibinfo{pages}{023}
  (\bibinfo{year}{2017}), \eprint{1611.07975}.

\bibitem[{\citenamefont{Csaki et~al.}(2017)\citenamefont{Csaki, Kuflik, and
  Lombardo}}]{Csaki:2017spo}
\bibinfo{author}{\bibfnamefont{C.}~\bibnamefont{Csaki}},
  \bibinfo{author}{\bibfnamefont{E.}~\bibnamefont{Kuflik}}, \bibnamefont{and}
  \bibinfo{author}{\bibfnamefont{S.}~\bibnamefont{Lombardo}},
  \bibinfo{journal}{Phys. Rev. D} \textbf{\bibinfo{volume}{96}},
  \bibinfo{pages}{055013} (\bibinfo{year}{2017}), \eprint{1703.06884}.

\bibitem[{\citenamefont{Chacko et~al.}(2018)\citenamefont{Chacko, Curtin,
  Geller, and Tsai}}]{Chacko:2018vss}
\bibinfo{author}{\bibfnamefont{Z.}~\bibnamefont{Chacko}},
  \bibinfo{author}{\bibfnamefont{D.}~\bibnamefont{Curtin}},
  \bibinfo{author}{\bibfnamefont{M.}~\bibnamefont{Geller}}, \bibnamefont{and}
  \bibinfo{author}{\bibfnamefont{Y.}~\bibnamefont{Tsai}},
  \bibinfo{journal}{JHEP} \textbf{\bibinfo{volume}{09}}, \bibinfo{pages}{163}
  (\bibinfo{year}{2018}), \eprint{1803.03263}.

\bibitem[{\citenamefont{Elor et~al.}(2018)\citenamefont{Elor, Liu, Slatyer, and
  Soreq}}]{Elor:2018xku}
\bibinfo{author}{\bibfnamefont{G.}~\bibnamefont{Elor}},
  \bibinfo{author}{\bibfnamefont{H.}~\bibnamefont{Liu}},
  \bibinfo{author}{\bibfnamefont{T.~R.} \bibnamefont{Slatyer}},
  \bibnamefont{and} \bibinfo{author}{\bibfnamefont{Y.}~\bibnamefont{Soreq}},
  \bibinfo{journal}{Phys. Rev. D} \textbf{\bibinfo{volume}{98}},
  \bibinfo{pages}{036015} (\bibinfo{year}{2018}), \eprint{1801.07723}.

\bibitem[{\citenamefont{Hochberg et~al.}(2019)\citenamefont{Hochberg, Kuflik,
  and Murayama}}]{Hochberg:2018vdo}
\bibinfo{author}{\bibfnamefont{Y.}~\bibnamefont{Hochberg}},
  \bibinfo{author}{\bibfnamefont{E.}~\bibnamefont{Kuflik}}, \bibnamefont{and}
  \bibinfo{author}{\bibfnamefont{H.}~\bibnamefont{Murayama}},
  \bibinfo{journal}{Phys. Rev. D} \textbf{\bibinfo{volume}{99}},
  \bibinfo{pages}{015005} (\bibinfo{year}{2019}), \eprint{1805.09345}.

\bibitem[{\citenamefont{Francis et~al.}(2018)\citenamefont{Francis, Hudspith,
  Lewis, and Tulin}}]{Francis:2018xjd}
\bibinfo{author}{\bibfnamefont{A.}~\bibnamefont{Francis}},
  \bibinfo{author}{\bibfnamefont{R.~J.} \bibnamefont{Hudspith}},
  \bibinfo{author}{\bibfnamefont{R.}~\bibnamefont{Lewis}}, \bibnamefont{and}
  \bibinfo{author}{\bibfnamefont{S.}~\bibnamefont{Tulin}},
  \bibinfo{journal}{JHEP} \textbf{\bibinfo{volume}{12}}, \bibinfo{pages}{118}
  (\bibinfo{year}{2018}), \eprint{1809.09117}.

\bibitem[{\citenamefont{Harigaya et~al.}(2020)\citenamefont{Harigaya, Mcgehee,
  Murayama, and Schutz}}]{Harigaya:2019shz}
\bibinfo{author}{\bibfnamefont{K.}~\bibnamefont{Harigaya}},
  \bibinfo{author}{\bibfnamefont{R.}~\bibnamefont{Mcgehee}},
  \bibinfo{author}{\bibfnamefont{H.}~\bibnamefont{Murayama}}, \bibnamefont{and}
  \bibinfo{author}{\bibfnamefont{K.}~\bibnamefont{Schutz}},
  \bibinfo{journal}{JHEP} \textbf{\bibinfo{volume}{05}}, \bibinfo{pages}{155}
  (\bibinfo{year}{2020}), \eprint{1905.08798}.

\bibitem[{\citenamefont{Ibe et~al.}(2019)\citenamefont{Ibe, Kamada, Kobayashi,
  Kuwahara, and Nakano}}]{Ibe:2019ena}
\bibinfo{author}{\bibfnamefont{M.}~\bibnamefont{Ibe}},
  \bibinfo{author}{\bibfnamefont{A.}~\bibnamefont{Kamada}},
  \bibinfo{author}{\bibfnamefont{S.}~\bibnamefont{Kobayashi}},
  \bibinfo{author}{\bibfnamefont{T.}~\bibnamefont{Kuwahara}}, \bibnamefont{and}
  \bibinfo{author}{\bibfnamefont{W.}~\bibnamefont{Nakano}},
  \bibinfo{journal}{Phys. Rev. D} \textbf{\bibinfo{volume}{100}},
  \bibinfo{pages}{075022} (\bibinfo{year}{2019}), \eprint{1907.03404}.

\bibitem[{\citenamefont{Dunsky et~al.}(2020)\citenamefont{Dunsky, Hall, and
  Harigaya}}]{Dunsky:2019upk}
\bibinfo{author}{\bibfnamefont{D.}~\bibnamefont{Dunsky}},
  \bibinfo{author}{\bibfnamefont{L.~J.} \bibnamefont{Hall}}, \bibnamefont{and}
  \bibinfo{author}{\bibfnamefont{K.}~\bibnamefont{Harigaya}},
  \bibinfo{journal}{JHEP} \textbf{\bibinfo{volume}{02}}, \bibinfo{pages}{078}
  (\bibinfo{year}{2020}), \eprint{1908.02756}.

\bibitem[{\citenamefont{Cs\'aki et~al.}(2020)\citenamefont{Cs\'aki, Guan, Ma,
  and Shu}}]{Csaki:2019qgb}
\bibinfo{author}{\bibfnamefont{C.}~\bibnamefont{Cs\'aki}},
  \bibinfo{author}{\bibfnamefont{C.-S.} \bibnamefont{Guan}},
  \bibinfo{author}{\bibfnamefont{T.}~\bibnamefont{Ma}}, \bibnamefont{and}
  \bibinfo{author}{\bibfnamefont{J.}~\bibnamefont{Shu}},
  \bibinfo{journal}{JHEP} \textbf{\bibinfo{volume}{12}}, \bibinfo{pages}{005}
  (\bibinfo{year}{2020}), \eprint{1910.14085}.

\bibitem[{\citenamefont{Koren and McGehee}(2020)}]{Koren:2019iuv}
\bibinfo{author}{\bibfnamefont{S.}~\bibnamefont{Koren}} \bibnamefont{and}
  \bibinfo{author}{\bibfnamefont{R.}~\bibnamefont{McGehee}},
  \bibinfo{journal}{Phys. Rev. D} \textbf{\bibinfo{volume}{101}},
  \bibinfo{pages}{055024} (\bibinfo{year}{2020}), \eprint{1908.03559}.

\bibitem[{\citenamefont{Terning et~al.}(2019)\citenamefont{Terning, Verhaaren,
  and Zora}}]{Terning:2019hgj}
\bibinfo{author}{\bibfnamefont{J.}~\bibnamefont{Terning}},
  \bibinfo{author}{\bibfnamefont{C.~B.} \bibnamefont{Verhaaren}},
  \bibnamefont{and} \bibinfo{author}{\bibfnamefont{K.}~\bibnamefont{Zora}},
  \bibinfo{journal}{Phys. Rev. D} \textbf{\bibinfo{volume}{99}},
  \bibinfo{pages}{095020} (\bibinfo{year}{2019}), \eprint{1902.08211}.

\bibitem[{\citenamefont{Johns and Koren}(2020)}]{Johns:2020rtp}
\bibinfo{author}{\bibfnamefont{L.}~\bibnamefont{Johns}} \bibnamefont{and}
  \bibinfo{author}{\bibfnamefont{S.}~\bibnamefont{Koren}}
  (\bibinfo{year}{2020}), \eprint{2012.06591}.

\bibitem[{\citenamefont{Roux and Cline}(2020)}]{Roux:2020wkp}
\bibinfo{author}{\bibfnamefont{J.-S.} \bibnamefont{Roux}} \bibnamefont{and}
  \bibinfo{author}{\bibfnamefont{J.~M.} \bibnamefont{Cline}},
  \bibinfo{journal}{Phys. Rev. D} \textbf{\bibinfo{volume}{102}},
  \bibinfo{pages}{063518} (\bibinfo{year}{2020}), \eprint{2001.11504}.

\bibitem[{\citenamefont{Ritter and Volkas}(2021)}]{Ritter:2021hgu}
\bibinfo{author}{\bibfnamefont{A.~C.} \bibnamefont{Ritter}} \bibnamefont{and}
  \bibinfo{author}{\bibfnamefont{R.~R.} \bibnamefont{Volkas}}
  (\bibinfo{year}{2021}), \eprint{2101.07421}.

\bibitem[{\citenamefont{Curtin and Gryba}(2021)}]{Curtin:2021alk}
\bibinfo{author}{\bibfnamefont{D.}~\bibnamefont{Curtin}} \bibnamefont{and}
  \bibinfo{author}{\bibfnamefont{S.}~\bibnamefont{Gryba}}
  (\bibinfo{year}{2021}), \eprint{2101.11019}.

\bibitem[{\citenamefont{Curtin et~al.}(2021)\citenamefont{Curtin, Gryba,
  Setford, Hooper, and Scholtz}}]{Curtin:2021spx}
\bibinfo{author}{\bibfnamefont{D.}~\bibnamefont{Curtin}},
  \bibinfo{author}{\bibfnamefont{S.}~\bibnamefont{Gryba}},
  \bibinfo{author}{\bibfnamefont{J.}~\bibnamefont{Setford}},
  \bibinfo{author}{\bibfnamefont{D.}~\bibnamefont{Hooper}}, \bibnamefont{and}
  \bibinfo{author}{\bibfnamefont{J.}~\bibnamefont{Scholtz}}
  (\bibinfo{year}{2021}), \eprint{2106.12578}.

\bibitem[{\citenamefont{Fields et~al.}(2020)\citenamefont{Fields, Olive, Yeh,
  and Young}}]{Fields:2019pfx}
\bibinfo{author}{\bibfnamefont{B.~D.} \bibnamefont{Fields}},
  \bibinfo{author}{\bibfnamefont{K.~A.} \bibnamefont{Olive}},
  \bibinfo{author}{\bibfnamefont{T.-H.} \bibnamefont{Yeh}}, \bibnamefont{and}
  \bibinfo{author}{\bibfnamefont{C.}~\bibnamefont{Young}},
  \bibinfo{journal}{JCAP} \textbf{\bibinfo{volume}{03}}, \bibinfo{pages}{010}
  (\bibinfo{year}{2020}), \bibinfo{note}{[Erratum: JCAP 11, E02 (2020)]},
  \eprint{1912.01132}.

\bibitem[{Note3()}]{Note3}
Note3, \bibinfo{note}{we could drop this caveat by extending the transformation
  further so that the atomic reaction rates scale appropriately as well. Doing
  so would also extend the challenge (significantly) of finding a model for
  which a direction in the parameter space corresponds to the scaling
  transformation.}

\bibitem[{Note4()}]{Note4}
Note4, \bibinfo{note}{the very mild symmetry breaking from non-zero neutrino
  masses could be removed by also scaling them by a factor of $\lambda $.}

\bibitem[{Note5()}]{Note5}
Note5, \bibinfo{note}{we include this caveat because we do not have a proof
  that the symmetry holds in the full non-linear theory. It might. We do know
  it holds to second order in perturbations; {\protect \bf see our
  Supplementary Material, which includes \protect \citep {Ma:1995ey, Zahn_2003,
  Bartolo:2007ax}}.}

\bibitem[{\citenamefont{Fixsen et~al.}(1996)\citenamefont{Fixsen, Cheng, Gales,
  Mather, Shafer, and Wright}}]{Fixsen1996}
\bibinfo{author}{\bibfnamefont{D.~J.} \bibnamefont{Fixsen}},
  \bibinfo{author}{\bibfnamefont{E.~S.} \bibnamefont{Cheng}},
  \bibinfo{author}{\bibfnamefont{J.~M.} \bibnamefont{Gales}},
  \bibinfo{author}{\bibfnamefont{J.~C.} \bibnamefont{Mather}},
  \bibinfo{author}{\bibfnamefont{R.~A.} \bibnamefont{Shafer}},
  \bibnamefont{and} \bibinfo{author}{\bibfnamefont{E.~L.}
  \bibnamefont{Wright}}, \bibinfo{journal}{Astrophys. J.}
  \textbf{\bibinfo{volume}{473}}, \bibinfo{pages}{576} (\bibinfo{year}{1996}),
  \eprint{astro-ph/9605054}.

\bibitem[{\citenamefont{Fixsen}(2009)}]{fixsen09}
\bibinfo{author}{\bibfnamefont{D.}~\bibnamefont{Fixsen}}, \bibinfo{journal}{The
  Astrophysical Journal} \textbf{\bibinfo{volume}{707}}, \bibinfo{pages}{916}
  (\bibinfo{year}{2009}).

\bibitem[{Note6()}]{Note6}
Note6, \bibinfo{note}{a similar point was made recently in Ref.~\cite
  {ivanov2020h}}.

\bibitem[{\citenamefont{Blinnikov and Khlopov}(1983)}]{Blinnikov:1983gh}
\bibinfo{author}{\bibfnamefont{S.~I.} \bibnamefont{Blinnikov}}
  \bibnamefont{and} \bibinfo{author}{\bibfnamefont{M.}~\bibnamefont{Khlopov}},
  \bibinfo{journal}{Sov. Astron.} \textbf{\bibinfo{volume}{27}},
  \bibinfo{pages}{371} (\bibinfo{year}{1983}).

\bibitem[{\citenamefont{Ackerman et~al.}(2009)\citenamefont{Ackerman, Buckley,
  Carroll, and Kamionkowski}}]{Ackerman:2008gi}
\bibinfo{author}{\bibfnamefont{L.}~\bibnamefont{Ackerman}},
  \bibinfo{author}{\bibfnamefont{M.~R.} \bibnamefont{Buckley}},
  \bibinfo{author}{\bibfnamefont{S.~M.} \bibnamefont{Carroll}},
  \bibnamefont{and}
  \bibinfo{author}{\bibfnamefont{M.}~\bibnamefont{Kamionkowski}},
  \bibinfo{journal}{Phys. Rev. D} \textbf{\bibinfo{volume}{79}},
  \bibinfo{pages}{023519} (\bibinfo{year}{2009}), \eprint{0810.5126}.

\bibitem[{\citenamefont{Feng et~al.}(2009)\citenamefont{Feng, Kaplinghat, Tu,
  and Yu}}]{Feng:2009mn}
\bibinfo{author}{\bibfnamefont{J.~L.} \bibnamefont{Feng}},
  \bibinfo{author}{\bibfnamefont{M.}~\bibnamefont{Kaplinghat}},
  \bibinfo{author}{\bibfnamefont{H.}~\bibnamefont{Tu}}, \bibnamefont{and}
  \bibinfo{author}{\bibfnamefont{H.-B.} \bibnamefont{Yu}},
  \bibinfo{journal}{JCAP} \textbf{\bibinfo{volume}{0907}}, \bibinfo{pages}{004}
  (\bibinfo{year}{2009}), \eprint{0905.3039}.

\bibitem[{\citenamefont{Agrawal et~al.}(2017)\citenamefont{Agrawal, Cyr-Racine,
  Randall, and Scholtz}}]{Agrawal:2016quu}
\bibinfo{author}{\bibfnamefont{P.}~\bibnamefont{Agrawal}},
  \bibinfo{author}{\bibfnamefont{F.-Y.} \bibnamefont{Cyr-Racine}},
  \bibinfo{author}{\bibfnamefont{L.}~\bibnamefont{Randall}}, \bibnamefont{and}
  \bibinfo{author}{\bibfnamefont{J.}~\bibnamefont{Scholtz}},
  \bibinfo{journal}{JCAP} \textbf{\bibinfo{volume}{1705}}, \bibinfo{pages}{022}
  (\bibinfo{year}{2017}), \eprint{1610.04611}.

\bibitem[{\citenamefont{Foot and Mitra}(2002)}]{Foot:2002iy}
\bibinfo{author}{\bibfnamefont{R.}~\bibnamefont{Foot}} \bibnamefont{and}
  \bibinfo{author}{\bibfnamefont{S.}~\bibnamefont{Mitra}},
  \bibinfo{journal}{Phys. Rev.} \textbf{\bibinfo{volume}{D66}},
  \bibinfo{pages}{061301} (\bibinfo{year}{2002}), \eprint{hep-ph/0204256}.

\bibitem[{\citenamefont{Foot and Volkas}(2003)}]{Foot:2003jt}
\bibinfo{author}{\bibfnamefont{R.}~\bibnamefont{Foot}} \bibnamefont{and}
  \bibinfo{author}{\bibfnamefont{R.~R.} \bibnamefont{Volkas}},
  \bibinfo{journal}{Phys. Rev.} \textbf{\bibinfo{volume}{D68}},
  \bibinfo{pages}{021304} (\bibinfo{year}{2003}), \eprint{hep-ph/0304261}.

\bibitem[{\citenamefont{Foot}(2004)}]{Foot:2004pa}
\bibinfo{author}{\bibfnamefont{R.}~\bibnamefont{Foot}}, \bibinfo{journal}{Int.
  J. Mod. Phys.} \textbf{\bibinfo{volume}{D13}}, \bibinfo{pages}{2161}
  (\bibinfo{year}{2004}), \eprint{astro-ph/0407623}.

\bibitem[{\citenamefont{Foot and Volkas}(2004)}]{Foot:2004wz}
\bibinfo{author}{\bibfnamefont{R.}~\bibnamefont{Foot}} \bibnamefont{and}
  \bibinfo{author}{\bibfnamefont{R.~R.} \bibnamefont{Volkas}},
  \bibinfo{journal}{Phys. Rev.} \textbf{\bibinfo{volume}{D70}},
  \bibinfo{pages}{123508} (\bibinfo{year}{2004}), \eprint{astro-ph/0407522}.

\bibitem[{\citenamefont{Foot et~al.}(2008)\citenamefont{Foot, Kobakhidze,
  McDonald, and Volkas}}]{Foot:2007iy}
\bibinfo{author}{\bibfnamefont{R.}~\bibnamefont{Foot}},
  \bibinfo{author}{\bibfnamefont{A.}~\bibnamefont{Kobakhidze}},
  \bibinfo{author}{\bibfnamefont{K.~L.} \bibnamefont{McDonald}},
  \bibnamefont{and} \bibinfo{author}{\bibfnamefont{R.~R.}
  \bibnamefont{Volkas}}, \bibinfo{journal}{Phys. Rev.}
  \textbf{\bibinfo{volume}{D77}}, \bibinfo{pages}{035006}
  (\bibinfo{year}{2008}), \eprint{0709.2750}.

\bibitem[{\citenamefont{Foot}(2012)}]{Foot:2011ve}
\bibinfo{author}{\bibfnamefont{R.}~\bibnamefont{Foot}}, \bibinfo{journal}{Phys.
  Lett.} \textbf{\bibinfo{volume}{B711}}, \bibinfo{pages}{238}
  (\bibinfo{year}{2012}), \eprint{1111.6366}.

\bibitem[{\citenamefont{Foot}(2013)}]{Foot:2013vna}
\bibinfo{author}{\bibfnamefont{R.}~\bibnamefont{Foot}}, \bibinfo{journal}{Phys.
  Rev.} \textbf{\bibinfo{volume}{D88}}, \bibinfo{pages}{023520}
  (\bibinfo{year}{2013}), \eprint{1304.4717}.

\bibitem[{\citenamefont{Foot}(2014)}]{Foot:2014mia}
\bibinfo{author}{\bibfnamefont{R.}~\bibnamefont{Foot}}, \bibinfo{journal}{Int.
  J. Mod. Phys.} \textbf{\bibinfo{volume}{A29}}, \bibinfo{pages}{1430013}
  (\bibinfo{year}{2014}), \eprint{1401.3965}.

\bibitem[{\citenamefont{Foot and Vagnozzi}(2015)}]{Foot:2014uba}
\bibinfo{author}{\bibfnamefont{R.}~\bibnamefont{Foot}} \bibnamefont{and}
  \bibinfo{author}{\bibfnamefont{S.}~\bibnamefont{Vagnozzi}},
  \bibinfo{journal}{Phys. Rev.} \textbf{\bibinfo{volume}{D91}},
  \bibinfo{pages}{023512} (\bibinfo{year}{2015}), \eprint{1409.7174}.

\bibitem[{\citenamefont{Foot and Vagnozzi}(2016)}]{Foot:2016wvj}
\bibinfo{author}{\bibfnamefont{R.}~\bibnamefont{Foot}} \bibnamefont{and}
  \bibinfo{author}{\bibfnamefont{S.}~\bibnamefont{Vagnozzi}},
  \bibinfo{journal}{JCAP} \textbf{\bibinfo{volume}{1607}}, \bibinfo{pages}{013}
  (\bibinfo{year}{2016}), \eprint{1602.02467}.

\bibitem[{\citenamefont{Ciarcelluti}(2005{\natexlab{a}})}]{Ciarcelluti:2004ik}
\bibinfo{author}{\bibfnamefont{P.}~\bibnamefont{Ciarcelluti}},
  \bibinfo{journal}{Int. J. Mod. Phys. D} \textbf{\bibinfo{volume}{14}},
  \bibinfo{pages}{187} (\bibinfo{year}{2005}{\natexlab{a}}),
  \eprint{astro-ph/0409630}.

\bibitem[{\citenamefont{Ciarcelluti}(2005{\natexlab{b}})}]{Ciarcelluti:2004ip}
\bibinfo{author}{\bibfnamefont{P.}~\bibnamefont{Ciarcelluti}},
  \bibinfo{journal}{Int. J. Mod. Phys. D} \textbf{\bibinfo{volume}{14}},
  \bibinfo{pages}{223} (\bibinfo{year}{2005}{\natexlab{b}}),
  \eprint{astro-ph/0409633}.

\bibitem[{\citenamefont{Ciarcelluti and Lepidi}(2008)}]{Ciarcelluti:2008vs}
\bibinfo{author}{\bibfnamefont{P.}~\bibnamefont{Ciarcelluti}} \bibnamefont{and}
  \bibinfo{author}{\bibfnamefont{A.}~\bibnamefont{Lepidi}},
  \bibinfo{journal}{Phys. Rev. D} \textbf{\bibinfo{volume}{78}},
  \bibinfo{pages}{123003} (\bibinfo{year}{2008}), \eprint{0809.0677}.

\bibitem[{\citenamefont{Ciarcelluti}(2010)}]{Ciarcelluti:2010zz}
\bibinfo{author}{\bibfnamefont{P.}~\bibnamefont{Ciarcelluti}},
  \bibinfo{journal}{Int. J. Mod. Phys. D} \textbf{\bibinfo{volume}{19}},
  \bibinfo{pages}{2151} (\bibinfo{year}{2010}), \eprint{1102.5530}.

\bibitem[{\citenamefont{Ciarcelluti and
  Wallemacq}(2014{\natexlab{a}})}]{Ciarcelluti:2012zz}
\bibinfo{author}{\bibfnamefont{P.}~\bibnamefont{Ciarcelluti}} \bibnamefont{and}
  \bibinfo{author}{\bibfnamefont{Q.}~\bibnamefont{Wallemacq}},
  \bibinfo{journal}{Phys. Lett. B} \textbf{\bibinfo{volume}{729}},
  \bibinfo{pages}{62} (\bibinfo{year}{2014}{\natexlab{a}}), \eprint{1211.5354}.

\bibitem[{\citenamefont{Ciarcelluti and
  Wallemacq}(2014{\natexlab{b}})}]{Ciarcelluti:2014scd}
\bibinfo{author}{\bibfnamefont{P.}~\bibnamefont{Ciarcelluti}} \bibnamefont{and}
  \bibinfo{author}{\bibfnamefont{Q.}~\bibnamefont{Wallemacq}},
  \bibinfo{journal}{Adv. High Energy Phys.} \textbf{\bibinfo{volume}{2014}},
  \bibinfo{pages}{148319} (\bibinfo{year}{2014}{\natexlab{b}}),
  \eprint{1401.4763}.

\bibitem[{\citenamefont{Cudell et~al.}(2014)\citenamefont{Cudell, Khlopov, and
  Wallemacq}}]{Cudell:2014wca}
\bibinfo{author}{\bibfnamefont{J.-R.} \bibnamefont{Cudell}},
  \bibinfo{author}{\bibfnamefont{M.~Y.} \bibnamefont{Khlopov}},
  \bibnamefont{and}
  \bibinfo{author}{\bibfnamefont{Q.}~\bibnamefont{Wallemacq}},
  \bibinfo{journal}{Mod. Phys. Lett. A} \textbf{\bibinfo{volume}{29}},
  \bibinfo{pages}{1440006} (\bibinfo{year}{2014}), \eprint{1411.1655}.

\bibitem[{\citenamefont{Goldberg and Hall}(1986)}]{Goldberg:1986nk}
\bibinfo{author}{\bibfnamefont{H.}~\bibnamefont{Goldberg}} \bibnamefont{and}
  \bibinfo{author}{\bibfnamefont{L.~J.} \bibnamefont{Hall}},
  \bibinfo{journal}{Phys. Lett.} \textbf{\bibinfo{volume}{B174}},
  \bibinfo{pages}{151} (\bibinfo{year}{1986}).

\bibitem[{\citenamefont{Fargion et~al.}(2006)\citenamefont{Fargion, Khlopov,
  and Stephan}}]{Fargion:2005ep}
\bibinfo{author}{\bibfnamefont{D.}~\bibnamefont{Fargion}},
  \bibinfo{author}{\bibfnamefont{M.}~\bibnamefont{Khlopov}}, \bibnamefont{and}
  \bibinfo{author}{\bibfnamefont{C.~A.} \bibnamefont{Stephan}},
  \bibinfo{journal}{Class. Quant. Grav.} \textbf{\bibinfo{volume}{23}},
  \bibinfo{pages}{7305} (\bibinfo{year}{2006}), \eprint{astro-ph/0511789}.

\bibitem[{\citenamefont{Khlopov}(2006)}]{Khlopov:2005ew}
\bibinfo{author}{\bibfnamefont{M.~Y.} \bibnamefont{Khlopov}},
  \bibinfo{journal}{Pisma Zh. Eksp. Teor. Fiz.} \textbf{\bibinfo{volume}{83}},
  \bibinfo{pages}{3} (\bibinfo{year}{2006}), \eprint{astro-ph/0511796}.

\bibitem[{\citenamefont{Khlopov and Kouvaris}(2008)}]{Khlopov:2008ty}
\bibinfo{author}{\bibfnamefont{M.~Y.} \bibnamefont{Khlopov}} \bibnamefont{and}
  \bibinfo{author}{\bibfnamefont{C.}~\bibnamefont{Kouvaris}},
  \bibinfo{journal}{Phys. Rev. D} \textbf{\bibinfo{volume}{78}},
  \bibinfo{pages}{065040} (\bibinfo{year}{2008}), \eprint{0806.1191}.

\bibitem[{\citenamefont{Kaplan et~al.}(2010)\citenamefont{Kaplan, Krnjaic,
  Rehermann, and Wells}}]{Kaplan:2009de}
\bibinfo{author}{\bibfnamefont{D.~E.} \bibnamefont{Kaplan}},
  \bibinfo{author}{\bibfnamefont{G.~Z.} \bibnamefont{Krnjaic}},
  \bibinfo{author}{\bibfnamefont{K.~R.} \bibnamefont{Rehermann}},
  \bibnamefont{and} \bibinfo{author}{\bibfnamefont{C.~M.} \bibnamefont{Wells}},
  \bibinfo{journal}{\jcap} \textbf{\bibinfo{volume}{1005}},
  \bibinfo{pages}{021} (\bibinfo{year}{2010}), \eprint{0909.0753}.

\bibitem[{\citenamefont{Khlopov et~al.}(2010)\citenamefont{Khlopov, Mayorov,
  and Soldatov}}]{Khlopov:2010pq}
\bibinfo{author}{\bibfnamefont{M.~Y.} \bibnamefont{Khlopov}},
  \bibinfo{author}{\bibfnamefont{A.~G.} \bibnamefont{Mayorov}},
  \bibnamefont{and} \bibinfo{author}{\bibfnamefont{E.~Y.}
  \bibnamefont{Soldatov}}, \bibinfo{journal}{Int. J. Mod. Phys. D}
  \textbf{\bibinfo{volume}{19}}, \bibinfo{pages}{1385} (\bibinfo{year}{2010}),
  \eprint{1003.1144}.

\bibitem[{\citenamefont{Kaplan et~al.}(2011)\citenamefont{Kaplan, Krnjaic,
  Rehermann, and Wells}}]{Kaplan:2011yj}
\bibinfo{author}{\bibfnamefont{D.~E.} \bibnamefont{Kaplan}},
  \bibinfo{author}{\bibfnamefont{G.~Z.} \bibnamefont{Krnjaic}},
  \bibinfo{author}{\bibfnamefont{K.~R.} \bibnamefont{Rehermann}},
  \bibnamefont{and} \bibinfo{author}{\bibfnamefont{C.~M.} \bibnamefont{Wells}},
  \bibinfo{journal}{\jcap} \textbf{\bibinfo{volume}{1110}},
  \bibinfo{pages}{011} (\bibinfo{year}{2011}), \eprint{1105.2073}.

\bibitem[{\citenamefont{Khlopov}(2011)}]{Khlopov:2011tn}
\bibinfo{author}{\bibfnamefont{M.~Y.} \bibnamefont{Khlopov}},
  \bibinfo{journal}{Mod. Phys. Lett. A} \textbf{\bibinfo{volume}{26}},
  \bibinfo{pages}{2823} (\bibinfo{year}{2011}), \eprint{1111.2838}.

\bibitem[{\citenamefont{Behbahani et~al.}(2011)\citenamefont{Behbahani,
  Jankowiak, Rube, and Wacker}}]{Behbahani:2010xa}
\bibinfo{author}{\bibfnamefont{S.~R.} \bibnamefont{Behbahani}},
  \bibinfo{author}{\bibfnamefont{M.}~\bibnamefont{Jankowiak}},
  \bibinfo{author}{\bibfnamefont{T.}~\bibnamefont{Rube}}, \bibnamefont{and}
  \bibinfo{author}{\bibfnamefont{J.~G.} \bibnamefont{Wacker}},
  \bibinfo{journal}{Adv. High Energy Phys.} \textbf{\bibinfo{volume}{2011}},
  \bibinfo{pages}{709492} (\bibinfo{year}{2011}), \eprint{1009.3523}.

\bibitem[{\citenamefont{Cline et~al.}(2012)\citenamefont{Cline, Liu, and
  Xue}}]{Cline:2012is}
\bibinfo{author}{\bibfnamefont{J.~M.} \bibnamefont{Cline}},
  \bibinfo{author}{\bibfnamefont{Z.}~\bibnamefont{Liu}}, \bibnamefont{and}
  \bibinfo{author}{\bibfnamefont{W.}~\bibnamefont{Xue}},
  \bibinfo{journal}{Phys. Rev. D} \textbf{\bibinfo{volume}{85}},
  \bibinfo{pages}{101302} (\bibinfo{year}{2012}), \eprint{1201.4858}.

\bibitem[{\citenamefont{Cyr-Racine and Sigurdson}(2013)}]{Cyr-Racine:2013ab}
\bibinfo{author}{\bibfnamefont{F.-Y.} \bibnamefont{Cyr-Racine}}
  \bibnamefont{and}
  \bibinfo{author}{\bibfnamefont{K.}~\bibnamefont{Sigurdson}},
  \bibinfo{journal}{Phys. Rev. D} \textbf{\bibinfo{volume}{87}},
  \bibinfo{pages}{103515} (\bibinfo{year}{2013}), \eprint{1209.5752}.

\bibitem[{\citenamefont{Cline et~al.}(2014)\citenamefont{Cline, Liu, Moore, and
  Xue}}]{Cline:2013pca}
\bibinfo{author}{\bibfnamefont{J.~M.} \bibnamefont{Cline}},
  \bibinfo{author}{\bibfnamefont{Z.}~\bibnamefont{Liu}},
  \bibinfo{author}{\bibfnamefont{G.}~\bibnamefont{Moore}}, \bibnamefont{and}
  \bibinfo{author}{\bibfnamefont{W.}~\bibnamefont{Xue}},
  \bibinfo{journal}{Phys. Rev. D} \textbf{\bibinfo{volume}{89}},
  \bibinfo{pages}{043514} (\bibinfo{year}{2014}), \eprint{1311.6468}.

\bibitem[{\citenamefont{Cyr-Racine et~al.}(2014)\citenamefont{Cyr-Racine,
  de~Putter, Raccanelli, and Sigurdson}}]{Cyr-Racine:2013fsa}
\bibinfo{author}{\bibfnamefont{F.-Y.} \bibnamefont{Cyr-Racine}},
  \bibinfo{author}{\bibfnamefont{R.}~\bibnamefont{de~Putter}},
  \bibinfo{author}{\bibfnamefont{A.}~\bibnamefont{Raccanelli}},
  \bibnamefont{and}
  \bibinfo{author}{\bibfnamefont{K.}~\bibnamefont{Sigurdson}},
  \bibinfo{journal}{Phys. Rev. D} \textbf{\bibinfo{volume}{89}},
  \bibinfo{pages}{063517} (\bibinfo{year}{2014}), \eprint{1310.3278}.

\bibitem[{\citenamefont{Fan et~al.}(2013{\natexlab{a}})\citenamefont{Fan, Katz,
  Randall, and Reece}}]{Fan:2013tia}
\bibinfo{author}{\bibfnamefont{J.~J.} \bibnamefont{Fan}},
  \bibinfo{author}{\bibfnamefont{A.}~\bibnamefont{Katz}},
  \bibinfo{author}{\bibfnamefont{L.}~\bibnamefont{Randall}}, \bibnamefont{and}
  \bibinfo{author}{\bibfnamefont{M.}~\bibnamefont{Reece}},
  \bibinfo{journal}{Phys. Rev. Lett.} \textbf{\bibinfo{volume}{110}},
  \bibinfo{pages}{211302} (\bibinfo{year}{2013}{\natexlab{a}}),
  \eprint{1303.3271}.

\bibitem[{\citenamefont{Fan et~al.}(2013{\natexlab{b}})\citenamefont{Fan, Katz,
  Randall, and Reece}}]{Fan:2013yva}
\bibinfo{author}{\bibfnamefont{J.}~\bibnamefont{Fan}},
  \bibinfo{author}{\bibfnamefont{A.}~\bibnamefont{Katz}},
  \bibinfo{author}{\bibfnamefont{L.}~\bibnamefont{Randall}}, \bibnamefont{and}
  \bibinfo{author}{\bibfnamefont{M.}~\bibnamefont{Reece}},
  \bibinfo{journal}{Phys. Dark Univ.} \textbf{\bibinfo{volume}{2}},
  \bibinfo{pages}{139} (\bibinfo{year}{2013}{\natexlab{b}}),
  \eprint{1303.1521}.

\bibitem[{\citenamefont{McCullough and Randall}(2013)}]{McCullough:2013jma}
\bibinfo{author}{\bibfnamefont{M.}~\bibnamefont{McCullough}} \bibnamefont{and}
  \bibinfo{author}{\bibfnamefont{L.}~\bibnamefont{Randall}},
  \bibinfo{journal}{\jcap} \textbf{\bibinfo{volume}{1310}},
  \bibinfo{pages}{058} (\bibinfo{year}{2013}), \eprint{1307.4095}.

\bibitem[{\citenamefont{Randall and Scholtz}(2015)}]{Randall:2014kta}
\bibinfo{author}{\bibfnamefont{L.}~\bibnamefont{Randall}} \bibnamefont{and}
  \bibinfo{author}{\bibfnamefont{J.}~\bibnamefont{Scholtz}},
  \bibinfo{journal}{\jcap} \textbf{\bibinfo{volume}{1509}},
  \bibinfo{pages}{057} (\bibinfo{year}{2015}), \eprint{1412.1839}.

\bibitem[{\citenamefont{Khlopov}(2014)}]{Khlopov:2014bia}
\bibinfo{author}{\bibfnamefont{M.~Y.} \bibnamefont{Khlopov}},
  \bibinfo{journal}{Int. J. Mod. Phys. A} \textbf{\bibinfo{volume}{29}},
  \bibinfo{pages}{1443002} (\bibinfo{year}{2014}), \eprint{1402.0181}.

\bibitem[{\citenamefont{Pearce et~al.}(2015)\citenamefont{Pearce, Petraki, and
  Kusenko}}]{Pearce:2015zca}
\bibinfo{author}{\bibfnamefont{L.}~\bibnamefont{Pearce}},
  \bibinfo{author}{\bibfnamefont{K.}~\bibnamefont{Petraki}}, \bibnamefont{and}
  \bibinfo{author}{\bibfnamefont{A.}~\bibnamefont{Kusenko}},
  \bibinfo{journal}{Phys. Rev. D} \textbf{\bibinfo{volume}{91}},
  \bibinfo{pages}{083532} (\bibinfo{year}{2015}), \eprint{1502.01755}.

\bibitem[{\citenamefont{Choquette and Cline}(2015)}]{Choquette:2015mca}
\bibinfo{author}{\bibfnamefont{J.}~\bibnamefont{Choquette}} \bibnamefont{and}
  \bibinfo{author}{\bibfnamefont{J.~M.} \bibnamefont{Cline}},
  \bibinfo{journal}{Phys. Rev. D} \textbf{\bibinfo{volume}{92}},
  \bibinfo{pages}{115011} (\bibinfo{year}{2015}), \eprint{1509.05764}.

\bibitem[{\citenamefont{Petraki et~al.}(2014)\citenamefont{Petraki, Pearce, and
  Kusenko}}]{Petraki:2014uza}
\bibinfo{author}{\bibfnamefont{K.}~\bibnamefont{Petraki}},
  \bibinfo{author}{\bibfnamefont{L.}~\bibnamefont{Pearce}}, \bibnamefont{and}
  \bibinfo{author}{\bibfnamefont{A.}~\bibnamefont{Kusenko}},
  \bibinfo{journal}{JCAP} \textbf{\bibinfo{volume}{07}}, \bibinfo{pages}{039}
  (\bibinfo{year}{2014}), \eprint{1403.1077}.

\bibitem[{\citenamefont{Cirelli et~al.}(2017)\citenamefont{Cirelli, Panci,
  Petraki, Sala, and Taoso}}]{Cirelli:2016rnw}
\bibinfo{author}{\bibfnamefont{M.}~\bibnamefont{Cirelli}},
  \bibinfo{author}{\bibfnamefont{P.}~\bibnamefont{Panci}},
  \bibinfo{author}{\bibfnamefont{K.}~\bibnamefont{Petraki}},
  \bibinfo{author}{\bibfnamefont{F.}~\bibnamefont{Sala}}, \bibnamefont{and}
  \bibinfo{author}{\bibfnamefont{M.}~\bibnamefont{Taoso}},
  \bibinfo{journal}{JCAP} \textbf{\bibinfo{volume}{05}}, \bibinfo{pages}{036}
  (\bibinfo{year}{2017}), \eprint{1612.07295}.

\bibitem[{\citenamefont{Petraki et~al.}(2017)\citenamefont{Petraki, Postma, and
  de~Vries}}]{Petraki:2016cnz}
\bibinfo{author}{\bibfnamefont{K.}~\bibnamefont{Petraki}},
  \bibinfo{author}{\bibfnamefont{M.}~\bibnamefont{Postma}}, \bibnamefont{and}
  \bibinfo{author}{\bibfnamefont{J.}~\bibnamefont{de~Vries}},
  \bibinfo{journal}{JHEP} \textbf{\bibinfo{volume}{04}}, \bibinfo{pages}{077}
  (\bibinfo{year}{2017}), \eprint{1611.01394}.

\bibitem[{\citenamefont{Curtin and Setford}(2021)}]{Curtin:2020tkm}
\bibinfo{author}{\bibfnamefont{D.}~\bibnamefont{Curtin}} \bibnamefont{and}
  \bibinfo{author}{\bibfnamefont{J.}~\bibnamefont{Setford}},
  \bibinfo{journal}{JHEP} \textbf{\bibinfo{volume}{03}}, \bibinfo{pages}{166}
  (\bibinfo{year}{2021}), \eprint{2010.00601}.

\bibitem[{Note7()}]{Note7}
Note7, \bibinfo{note}{we note that using the more recent estimate of
  $N_{\protect \rm eff}^{\protect \rm fs} = 3.044$ \cite {Froustey:2020mcq}
  would have minimal impact on our results.}

\bibitem[{\citenamefont{{Hou} et~al.}(2013)\citenamefont{{Hou}, {Keisler},
  {Knox}, {Millea}, and {Reichardt}}}]{Hou:2011ec}
\bibinfo{author}{\bibfnamefont{Z.}~\bibnamefont{{Hou}}},
  \bibinfo{author}{\bibfnamefont{R.}~\bibnamefont{{Keisler}}},
  \bibinfo{author}{\bibfnamefont{L.}~\bibnamefont{{Knox}}},
  \bibinfo{author}{\bibfnamefont{M.}~\bibnamefont{{Millea}}}, \bibnamefont{and}
  \bibinfo{author}{\bibfnamefont{C.}~\bibnamefont{{Reichardt}}},
  \bibinfo{journal}{\prd} \textbf{\bibinfo{volume}{87}}, \bibinfo{eid}{083008}
  (\bibinfo{year}{2013}), \eprint{1104.2333}.

\bibitem[{\citenamefont{Follin et~al.}(2015)\citenamefont{Follin, Knox, Millea,
  and Pan}}]{Follin:2015hya}
\bibinfo{author}{\bibfnamefont{B.}~\bibnamefont{Follin}},
  \bibinfo{author}{\bibfnamefont{L.}~\bibnamefont{Knox}},
  \bibinfo{author}{\bibfnamefont{M.}~\bibnamefont{Millea}}, \bibnamefont{and}
  \bibinfo{author}{\bibfnamefont{Z.}~\bibnamefont{Pan}},
  \bibinfo{journal}{Phys. Rev. Lett.} \textbf{\bibinfo{volume}{115}},
  \bibinfo{pages}{091301} (\bibinfo{year}{2015}), \eprint{1503.07863}.

\bibitem[{\citenamefont{Baumann et~al.}(2016)\citenamefont{Baumann, Green,
  Meyers, and Wallisch}}]{Baumann:2015rya}
\bibinfo{author}{\bibfnamefont{D.}~\bibnamefont{Baumann}},
  \bibinfo{author}{\bibfnamefont{D.}~\bibnamefont{Green}},
  \bibinfo{author}{\bibfnamefont{J.}~\bibnamefont{Meyers}}, \bibnamefont{and}
  \bibinfo{author}{\bibfnamefont{B.}~\bibnamefont{Wallisch}},
  \bibinfo{journal}{JCAP} \textbf{\bibinfo{volume}{01}}, \bibinfo{pages}{007}
  (\bibinfo{year}{2016}), \eprint{1508.06342}.

\bibitem[{\citenamefont{Aghanim et~al.}(2020{\natexlab{b}})}]{Planck:2019nip}
\bibinfo{author}{\bibfnamefont{N.}~\bibnamefont{Aghanim}} \bibnamefont{et~al.}
  (\bibinfo{collaboration}{Planck}), \bibinfo{journal}{Astron. Astrophys.}
  \textbf{\bibinfo{volume}{641}}, \bibinfo{pages}{A5}
  (\bibinfo{year}{2020}{\natexlab{b}}), \eprint{1907.12875}.

\bibitem[{\citenamefont{Beutler et~al.}(2011)\citenamefont{Beutler, Blake,
  Colless, Jones, Staveley-Smith, Campbell, Parker, Saunders, and
  Watson}}]{Beutler:2011}
\bibinfo{author}{\bibfnamefont{F.}~\bibnamefont{Beutler}},
  \bibinfo{author}{\bibfnamefont{C.}~\bibnamefont{Blake}},
  \bibinfo{author}{\bibfnamefont{M.}~\bibnamefont{Colless}},
  \bibinfo{author}{\bibfnamefont{D.~H.} \bibnamefont{Jones}},
  \bibinfo{author}{\bibfnamefont{L.}~\bibnamefont{Staveley-Smith}},
  \bibinfo{author}{\bibfnamefont{L.}~\bibnamefont{Campbell}},
  \bibinfo{author}{\bibfnamefont{Q.}~\bibnamefont{Parker}},
  \bibinfo{author}{\bibfnamefont{W.}~\bibnamefont{Saunders}}, \bibnamefont{and}
  \bibinfo{author}{\bibfnamefont{F.}~\bibnamefont{Watson}},
  \bibinfo{journal}{Monthly Notices of the Royal Astronomical Society}
  \textbf{\bibinfo{volume}{416}}, \bibinfo{pages}{3017} (\bibinfo{year}{2011}).

\bibitem[{\citenamefont{Ross et~al.}(2015)\citenamefont{Ross, Samushia,
  Howlett, Percival, Burden, and Manera}}]{Ross:2014qpa}
\bibinfo{author}{\bibfnamefont{A.~J.} \bibnamefont{Ross}},
  \bibinfo{author}{\bibfnamefont{L.}~\bibnamefont{Samushia}},
  \bibinfo{author}{\bibfnamefont{C.}~\bibnamefont{Howlett}},
  \bibinfo{author}{\bibfnamefont{W.~J.} \bibnamefont{Percival}},
  \bibinfo{author}{\bibfnamefont{A.}~\bibnamefont{Burden}}, \bibnamefont{and}
  \bibinfo{author}{\bibfnamefont{M.}~\bibnamefont{Manera}},
  \bibinfo{journal}{Mon. Not. Roy. Astron. Soc.}
  \textbf{\bibinfo{volume}{449}}, \bibinfo{pages}{835} (\bibinfo{year}{2015}),
  \eprint{1409.3242}.

\bibitem[{\citenamefont{Alam et~al.}(2017)}]{BOSS:2016wmc}
\bibinfo{author}{\bibfnamefont{S.}~\bibnamefont{Alam}} \bibnamefont{et~al.}
  (\bibinfo{collaboration}{BOSS}), \bibinfo{journal}{Mon. Not. Roy. Astron.
  Soc.} \textbf{\bibinfo{volume}{470}}, \bibinfo{pages}{2617}
  (\bibinfo{year}{2017}), \eprint{1607.03155}.

\bibitem[{\citenamefont{Seager et~al.}(1999)\citenamefont{Seager, Sasselov, and
  Scott}}]{Seager:1999bc}
\bibinfo{author}{\bibfnamefont{S.}~\bibnamefont{Seager}},
  \bibinfo{author}{\bibfnamefont{D.~D.} \bibnamefont{Sasselov}},
  \bibnamefont{and} \bibinfo{author}{\bibfnamefont{D.}~\bibnamefont{Scott}},
  \bibinfo{journal}{Astrophys. J. Lett.} \textbf{\bibinfo{volume}{523}},
  \bibinfo{pages}{L1} (\bibinfo{year}{1999}), \eprint{astro-ph/9909275}.

\bibitem[{\citenamefont{Wong et~al.}(2008)\citenamefont{Wong, Moss, and
  Scott}}]{Wong:2007ym}
\bibinfo{author}{\bibfnamefont{W.~Y.} \bibnamefont{Wong}},
  \bibinfo{author}{\bibfnamefont{A.}~\bibnamefont{Moss}}, \bibnamefont{and}
  \bibinfo{author}{\bibfnamefont{D.}~\bibnamefont{Scott}},
  \bibinfo{journal}{Mon. Not. Roy. Astron. Soc.}
  \textbf{\bibinfo{volume}{386}}, \bibinfo{pages}{1023} (\bibinfo{year}{2008}),
  \eprint{0711.1357}.

\bibitem[{\citenamefont{Esteban et~al.}(2020)\citenamefont{Esteban,
  Gonzalez-Garcia, Maltoni, Schwetz, and Zhou}}]{Esteban:2020cvm}
\bibinfo{author}{\bibfnamefont{I.}~\bibnamefont{Esteban}},
  \bibinfo{author}{\bibfnamefont{M.~C.} \bibnamefont{Gonzalez-Garcia}},
  \bibinfo{author}{\bibfnamefont{M.}~\bibnamefont{Maltoni}},
  \bibinfo{author}{\bibfnamefont{T.}~\bibnamefont{Schwetz}}, \bibnamefont{and}
  \bibinfo{author}{\bibfnamefont{A.}~\bibnamefont{Zhou}},
  \bibinfo{journal}{JHEP} \textbf{\bibinfo{volume}{09}}, \bibinfo{pages}{178}
  (\bibinfo{year}{2020}), \eprint{2007.14792}.

\bibitem[{\citenamefont{Lewis et~al.}(2000)\citenamefont{Lewis, Challinor, and
  Lasenby}}]{Lewis:1999camb}
\bibinfo{author}{\bibfnamefont{A.}~\bibnamefont{Lewis}},
  \bibinfo{author}{\bibfnamefont{A.}~\bibnamefont{Challinor}},
  \bibnamefont{and} \bibinfo{author}{\bibfnamefont{A.}~\bibnamefont{Lasenby}},
  \bibinfo{journal}{Astrophys. J.} \textbf{\bibinfo{volume}{538}},
  \bibinfo{pages}{473} (\bibinfo{year}{2000}), \eprint{astro-ph/9911177}.

\bibitem[{\citenamefont{Lewis and Bridle}(2002)}]{Lewis:2002mc}
\bibinfo{author}{\bibfnamefont{A.}~\bibnamefont{Lewis}} \bibnamefont{and}
  \bibinfo{author}{\bibfnamefont{S.}~\bibnamefont{Bridle}},
  \bibinfo{journal}{Phys. Rev. D} \textbf{\bibinfo{volume}{66}},
  \bibinfo{pages}{103511} (\bibinfo{year}{2002}), \eprint{astro-ph/0205436}.

\bibitem[{Note8()}]{Note8}
Note8, \bibinfo{note}{this prior is made necessary by our use of a mirror dark
  sector to mimic a $\Lambda $CDM model with scaled-up densities. It ensures
  that the energy densities of dark photons and dark atoms are always
  positive.}

\bibitem[{\citenamefont{Aver et~al.}(2021)\citenamefont{Aver, Berg, Olive,
  Pogge, Salzer, and Skillman}}]{aver2021}
\bibinfo{author}{\bibfnamefont{E.}~\bibnamefont{Aver}},
  \bibinfo{author}{\bibfnamefont{D.~A.} \bibnamefont{Berg}},
  \bibinfo{author}{\bibfnamefont{K.~A.} \bibnamefont{Olive}},
  \bibinfo{author}{\bibfnamefont{R.~W.} \bibnamefont{Pogge}},
  \bibinfo{author}{\bibfnamefont{J.~J.} \bibnamefont{Salzer}},
  \bibnamefont{and} \bibinfo{author}{\bibfnamefont{E.~D.}
  \bibnamefont{Skillman}}, \bibinfo{journal}{Journal of Cosmology and
  Astroparticle Physics} \textbf{\bibinfo{volume}{2021}}, \bibinfo{pages}{027}
  (\bibinfo{year}{2021}).

\bibitem[{\citenamefont{Consiglio et~al.}(2018)\citenamefont{Consiglio,
  de~Salas, Mangano, Miele, Pastor, and Pisanti}}]{Consiglio:2017pot}
\bibinfo{author}{\bibfnamefont{R.}~\bibnamefont{Consiglio}},
  \bibinfo{author}{\bibfnamefont{P.~F.} \bibnamefont{de~Salas}},
  \bibinfo{author}{\bibfnamefont{G.}~\bibnamefont{Mangano}},
  \bibinfo{author}{\bibfnamefont{G.}~\bibnamefont{Miele}},
  \bibinfo{author}{\bibfnamefont{S.}~\bibnamefont{Pastor}}, \bibnamefont{and}
  \bibinfo{author}{\bibfnamefont{O.}~\bibnamefont{Pisanti}},
  \bibinfo{journal}{Comput. Phys. Commun.} \textbf{\bibinfo{volume}{233}},
  \bibinfo{pages}{237} (\bibinfo{year}{2018}), \eprint{1712.04378}.

\bibitem[{Note9()}]{Note9}
Note9, \bibinfo{note}{j. Chluba, private communication}.

\bibitem[{\citenamefont{{Sekiguchi} and {Takahashi}}(2021)}]{Sekiguichi:2021}
\bibinfo{author}{\bibfnamefont{T.}~\bibnamefont{{Sekiguchi}}} \bibnamefont{and}
  \bibinfo{author}{\bibfnamefont{T.}~\bibnamefont{{Takahashi}}},
  \bibinfo{journal}{\prd} \textbf{\bibinfo{volume}{103}}, \bibinfo{eid}{083507}
  (\bibinfo{year}{2021}), \eprint{2007.03381}.

\bibitem[{\citenamefont{{Hart} and {Chluba}}(2020)}]{Hart:2020}
\bibinfo{author}{\bibfnamefont{L.}~\bibnamefont{{Hart}}} \bibnamefont{and}
  \bibinfo{author}{\bibfnamefont{J.}~\bibnamefont{{Chluba}}},
  \bibinfo{journal}{\mnras} \textbf{\bibinfo{volume}{493}},
  \bibinfo{pages}{3255} (\bibinfo{year}{2020}), \eprint{1912.03986}.

\bibitem[{\citenamefont{Burgess and Quevedo}(2021)}]{Burgess:2021qti}
\bibinfo{author}{\bibfnamefont{C.~P.} \bibnamefont{Burgess}} \bibnamefont{and}
  \bibinfo{author}{\bibfnamefont{F.}~\bibnamefont{Quevedo}}
  (\bibinfo{year}{2021}), \eprint{2110.10352}.

\bibitem[{\citenamefont{Burgess et~al.}(2021)\citenamefont{Burgess, Dineen, and
  Quevedo}}]{Burgess:2021obw}
\bibinfo{author}{\bibfnamefont{C.~P.} \bibnamefont{Burgess}},
  \bibinfo{author}{\bibfnamefont{D.}~\bibnamefont{Dineen}}, \bibnamefont{and}
  \bibinfo{author}{\bibfnamefont{F.}~\bibnamefont{Quevedo}}
  (\bibinfo{year}{2021}), \eprint{2111.07286}.

\bibitem[{\citenamefont{Aloni et~al.}(2021)\citenamefont{Aloni, Berlin, Joseph,
  Schmaltz, and Weiner}}]{Aloni:2021eaq}
\bibinfo{author}{\bibfnamefont{D.}~\bibnamefont{Aloni}},
  \bibinfo{author}{\bibfnamefont{A.}~\bibnamefont{Berlin}},
  \bibinfo{author}{\bibfnamefont{M.}~\bibnamefont{Joseph}},
  \bibinfo{author}{\bibfnamefont{M.}~\bibnamefont{Schmaltz}}, \bibnamefont{and}
  \bibinfo{author}{\bibfnamefont{N.}~\bibnamefont{Weiner}}
  (\bibinfo{year}{2021}), \eprint{2111.00014}.

\bibitem[{\citenamefont{Burdman et~al.}(2015)\citenamefont{Burdman, Chacko,
  Harnik, de~Lima, and Verhaaren}}]{Burdman:2014zta}
\bibinfo{author}{\bibfnamefont{G.}~\bibnamefont{Burdman}},
  \bibinfo{author}{\bibfnamefont{Z.}~\bibnamefont{Chacko}},
  \bibinfo{author}{\bibfnamefont{R.}~\bibnamefont{Harnik}},
  \bibinfo{author}{\bibfnamefont{L.}~\bibnamefont{de~Lima}}, \bibnamefont{and}
  \bibinfo{author}{\bibfnamefont{C.~B.} \bibnamefont{Verhaaren}},
  \bibinfo{journal}{Phys. Rev. D} \textbf{\bibinfo{volume}{91}},
  \bibinfo{pages}{055007} (\bibinfo{year}{2015}), \eprint{1411.3310}.

\bibitem[{\citenamefont{Blinov et~al.}(2021)\citenamefont{Blinov, Krnjaic, and
  Li}}]{Blinov:2021mdk}
\bibinfo{author}{\bibfnamefont{N.}~\bibnamefont{Blinov}},
  \bibinfo{author}{\bibfnamefont{G.}~\bibnamefont{Krnjaic}}, \bibnamefont{and}
  \bibinfo{author}{\bibfnamefont{S.~W.} \bibnamefont{Li}}
  (\bibinfo{year}{2021}), \eprint{2108.11386}.

\bibitem[{\citenamefont{Bansal et~al.}(2021)\citenamefont{Bansal, Kim, Kolda,
  Low, and Tsai}}]{Bansal:2021dfh}
\bibinfo{author}{\bibfnamefont{S.}~\bibnamefont{Bansal}},
  \bibinfo{author}{\bibfnamefont{J.~H.} \bibnamefont{Kim}},
  \bibinfo{author}{\bibfnamefont{C.}~\bibnamefont{Kolda}},
  \bibinfo{author}{\bibfnamefont{M.}~\bibnamefont{Low}}, \bibnamefont{and}
  \bibinfo{author}{\bibfnamefont{Y.}~\bibnamefont{Tsai}}
  (\bibinfo{year}{2021}), \eprint{2110.04317}.

\bibitem[{\citenamefont{Shandera et~al.}(2018)\citenamefont{Shandera, Jeong,
  and Gebhardt}}]{Shandera:2018xkn}
\bibinfo{author}{\bibfnamefont{S.}~\bibnamefont{Shandera}},
  \bibinfo{author}{\bibfnamefont{D.}~\bibnamefont{Jeong}}, \bibnamefont{and}
  \bibinfo{author}{\bibfnamefont{H.~S.~G.} \bibnamefont{Gebhardt}},
  \bibinfo{journal}{Phys. Rev. Lett.} \textbf{\bibinfo{volume}{120}},
  \bibinfo{pages}{241102} (\bibinfo{year}{2018}), \eprint{1802.08206}.

\bibitem[{\citenamefont{Singh et~al.}(2020)\citenamefont{Singh, Ryan, Magee,
  Akhter, Shandera, Jeong, and Hanna}}]{Singh:2020wiq}
\bibinfo{author}{\bibfnamefont{D.}~\bibnamefont{Singh}},
  \bibinfo{author}{\bibfnamefont{M.}~\bibnamefont{Ryan}},
  \bibinfo{author}{\bibfnamefont{R.}~\bibnamefont{Magee}},
  \bibinfo{author}{\bibfnamefont{T.}~\bibnamefont{Akhter}},
  \bibinfo{author}{\bibfnamefont{S.}~\bibnamefont{Shandera}},
  \bibinfo{author}{\bibfnamefont{D.}~\bibnamefont{Jeong}}, \bibnamefont{and}
  \bibinfo{author}{\bibfnamefont{C.}~\bibnamefont{Hanna}}
  (\bibinfo{year}{2020}), \eprint{2009.05209}.

\bibitem[{\citenamefont{Curtin and
  Setford}(2020{\natexlab{a}})}]{Curtin:2019lhm}
\bibinfo{author}{\bibfnamefont{D.}~\bibnamefont{Curtin}} \bibnamefont{and}
  \bibinfo{author}{\bibfnamefont{J.}~\bibnamefont{Setford}},
  \bibinfo{journal}{Phys. Lett. B} \textbf{\bibinfo{volume}{804}},
  \bibinfo{pages}{135391} (\bibinfo{year}{2020}{\natexlab{a}}),
  \eprint{1909.04071}.

\bibitem[{\citenamefont{Curtin and
  Setford}(2020{\natexlab{b}})}]{Curtin:2019ngc}
\bibinfo{author}{\bibfnamefont{D.}~\bibnamefont{Curtin}} \bibnamefont{and}
  \bibinfo{author}{\bibfnamefont{J.}~\bibnamefont{Setford}},
  \bibinfo{journal}{JHEP} \textbf{\bibinfo{volume}{03}}, \bibinfo{pages}{041}
  (\bibinfo{year}{2020}{\natexlab{b}}), \eprint{1909.04072}.

\bibitem[{\citenamefont{Agrawal and Randall}(2017)}]{Agrawal:2017pnb}
\bibinfo{author}{\bibfnamefont{P.}~\bibnamefont{Agrawal}} \bibnamefont{and}
  \bibinfo{author}{\bibfnamefont{L.}~\bibnamefont{Randall}},
  \bibinfo{journal}{JCAP} \textbf{\bibinfo{volume}{12}}, \bibinfo{pages}{019}
  (\bibinfo{year}{2017}), \eprint{1706.04195}.

\bibitem[{\citenamefont{Kramer and
  Randall}(2016{\natexlab{a}})}]{Kramer:2016dew}
\bibinfo{author}{\bibfnamefont{E.~D.} \bibnamefont{Kramer}} \bibnamefont{and}
  \bibinfo{author}{\bibfnamefont{L.}~\bibnamefont{Randall}},
  \bibinfo{journal}{Astrophys. J.} \textbf{\bibinfo{volume}{829}},
  \bibinfo{pages}{126} (\bibinfo{year}{2016}{\natexlab{a}}),
  \eprint{1603.03058}.

\bibitem[{\citenamefont{Kramer and
  Randall}(2016{\natexlab{b}})}]{Kramer:2016dqu}
\bibinfo{author}{\bibfnamefont{E.~D.} \bibnamefont{Kramer}} \bibnamefont{and}
  \bibinfo{author}{\bibfnamefont{L.}~\bibnamefont{Randall}},
  \bibinfo{journal}{Astrophys. J.} \textbf{\bibinfo{volume}{824}},
  \bibinfo{pages}{116} (\bibinfo{year}{2016}{\natexlab{b}}),
  \eprint{1604.01407}.

\bibitem[{\citenamefont{Schutz et~al.}(2018)\citenamefont{Schutz, Lin, Safdi,
  and Wu}}]{Schutz:2017tfp}
\bibinfo{author}{\bibfnamefont{K.}~\bibnamefont{Schutz}},
  \bibinfo{author}{\bibfnamefont{T.}~\bibnamefont{Lin}},
  \bibinfo{author}{\bibfnamefont{B.~R.} \bibnamefont{Safdi}}, \bibnamefont{and}
  \bibinfo{author}{\bibfnamefont{C.-L.} \bibnamefont{Wu}},
  \bibinfo{journal}{Phys. Rev. Lett.} \textbf{\bibinfo{volume}{121}},
  \bibinfo{pages}{081101} (\bibinfo{year}{2018}), \eprint{1711.03103}.

\bibitem[{\citenamefont{Buch et~al.}(2019)\citenamefont{Buch, Leung, and
  Fan}}]{Buch:2018qdr}
\bibinfo{author}{\bibfnamefont{J.}~\bibnamefont{Buch}},
  \bibinfo{author}{\bibfnamefont{S.~C.~J.} \bibnamefont{Leung}},
  \bibnamefont{and} \bibinfo{author}{\bibfnamefont{J.}~\bibnamefont{Fan}},
  \bibinfo{journal}{JCAP} \textbf{\bibinfo{volume}{04}}, \bibinfo{pages}{026}
  (\bibinfo{year}{2019}), \eprint{1808.05603}.

\bibitem[{\citenamefont{Dvali}(2010)}]{Dvali:2007hz}
\bibinfo{author}{\bibfnamefont{G.}~\bibnamefont{Dvali}},
  \bibinfo{journal}{Fortsch. Phys.} \textbf{\bibinfo{volume}{58}},
  \bibinfo{pages}{528} (\bibinfo{year}{2010}), \eprint{0706.2050}.

\bibitem[{\citenamefont{Dvali and Redi}(2009)}]{Dvali:2009ne}
\bibinfo{author}{\bibfnamefont{G.}~\bibnamefont{Dvali}} \bibnamefont{and}
  \bibinfo{author}{\bibfnamefont{M.}~\bibnamefont{Redi}},
  \bibinfo{journal}{Phys. Rev. D} \textbf{\bibinfo{volume}{80}},
  \bibinfo{pages}{055001} (\bibinfo{year}{2009}), \eprint{0905.1709}.

\bibitem[{\citenamefont{Arkani-Hamed et~al.}(2016)\citenamefont{Arkani-Hamed,
  Cohen, D'Agnolo, Hook, Kim, and Pinner}}]{Arkani-Hamed:2016rle}
\bibinfo{author}{\bibfnamefont{N.}~\bibnamefont{Arkani-Hamed}},
  \bibinfo{author}{\bibfnamefont{T.}~\bibnamefont{Cohen}},
  \bibinfo{author}{\bibfnamefont{R.~T.} \bibnamefont{D'Agnolo}},
  \bibinfo{author}{\bibfnamefont{A.}~\bibnamefont{Hook}},
  \bibinfo{author}{\bibfnamefont{H.~D.} \bibnamefont{Kim}}, \bibnamefont{and}
  \bibinfo{author}{\bibfnamefont{D.}~\bibnamefont{Pinner}},
  \bibinfo{journal}{Phys. Rev. Lett.} \textbf{\bibinfo{volume}{117}},
  \bibinfo{pages}{251801} (\bibinfo{year}{2016}), \eprint{1607.06821}.

\bibitem[{\citenamefont{Ma and Bertschinger}(1995)}]{Ma:1995ey}
\bibinfo{author}{\bibfnamefont{C.-P.} \bibnamefont{Ma}} \bibnamefont{and}
  \bibinfo{author}{\bibfnamefont{E.}~\bibnamefont{Bertschinger}},
  \bibinfo{journal}{Astrophys.J.} \textbf{\bibinfo{volume}{455}},
  \bibinfo{pages}{7} (\bibinfo{year}{1995}).

\bibitem[{\citenamefont{Bartolo et~al.}(2007)\citenamefont{Bartolo, Matarrese,
  and Riotto}}]{Bartolo:2007ax}
\bibinfo{author}{\bibfnamefont{N.}~\bibnamefont{Bartolo}},
  \bibinfo{author}{\bibfnamefont{S.}~\bibnamefont{Matarrese}},
  \bibnamefont{and} \bibinfo{author}{\bibfnamefont{A.}~\bibnamefont{Riotto}},
  in \emph{\bibinfo{booktitle}{{Les Houches Summer School - Session 86:
  Particle Physics and Cosmology: The Fabric of Spacetime}}}
  (\bibinfo{year}{2007}), \eprint{astro-ph/0703496}.

\bibitem[{\citenamefont{Ivanov et~al.}(2020{\natexlab{b}})\citenamefont{Ivanov,
  Ali-Ha{\"\i}moud, and Lesgourgues}}]{ivanov2020h}
\bibinfo{author}{\bibfnamefont{M.~M.} \bibnamefont{Ivanov}},
  \bibinfo{author}{\bibfnamefont{Y.}~\bibnamefont{Ali-Ha{\"\i}moud}},
  \bibnamefont{and}
  \bibinfo{author}{\bibfnamefont{J.}~\bibnamefont{Lesgourgues}},
  \bibinfo{journal}{Physical Review D} \textbf{\bibinfo{volume}{102}},
  \bibinfo{pages}{063515} (\bibinfo{year}{2020}{\natexlab{b}}).

\bibitem[{\citenamefont{Froustey et~al.}(2020)\citenamefont{Froustey, Pitrou,
  and Volpe}}]{Froustey:2020mcq}
\bibinfo{author}{\bibfnamefont{J.}~\bibnamefont{Froustey}},
  \bibinfo{author}{\bibfnamefont{C.}~\bibnamefont{Pitrou}}, \bibnamefont{and}
  \bibinfo{author}{\bibfnamefont{M.~C.} \bibnamefont{Volpe}},
  \bibinfo{journal}{JCAP} \textbf{\bibinfo{volume}{12}}, \bibinfo{pages}{015}
  (\bibinfo{year}{2020}), \eprint{2008.01074}.

\end{thebibliography}

\appendix
\section{Supplemental Material: Symmetry of the Equations of Motion}
In this supplementary material, we examine how the equations of motion governing the evolution of cosmological fluctuations are invariant under the scaling transformation introduced in the main text. We first consider the standard first-order equations, and then examine how the scaling transformation generalizes to higher order in perturbation theory.

\subsection{First-order equations}

As a starting point, let us first examine the  Boltzmann equations governing the evolution of photons and baryons fluctuations. Using the scale factor $a$ as our time variable, these take the form \citep{Ma:1995ey}
\begin{align}\label{eq:phot_bar}
   \frac{\pa F_{\gamma0}}{\pa a} & = - \frac{k }{ a^2 H}F_{\gamma 1} + 4\frac{\pa \phi}{\pa a} ,\\ 
    a^2 H\frac{\pa F_{\gamma1}}{\pa a} & = \frac{k}{3}(F_{\gamma0} - 2F_{\gamma2}) + \frac{4k}{3} \psi + \dot{\kappa}( \frac{4}{3}v_{\rm b}-F_{\gamma1}), \en
   a^2H\frac{\pa F_{\gamma2}}{\pa a} & = \frac{k}{5}(2F_{\gamma1} -3F_{\gamma3}) -\frac{9}{10}\dot{\kappa}F_{\gamma 2},\en
   a^2 H \frac{\pa F_{\gamma l}}{\pa a} & = \frac{k}{2l+1}\left[l F_{\gamma(l-1)} - (l+1)F_{\gamma(l+1)}\right] - \dot{\kappa}F_{\gamma l},\en
   \frac{\pa \delta_{\rm b}}{\pa a} & = - \frac{k}{a^2 H} v_{\rm b} + 3 \frac{\pa \phi}{\pa a}, \en
   a^2 H \frac{\pa v_{\rm b}}{\pa a} & = - a H v_{\rm b} + c_{\rm s}^2 k \delta_{\rm b} + \frac{\rho_\gamma }{\rho_{\rm b}} \dot{\kappa}(F_{\gamma1} -  \frac{4}{3}v_b),\nonumber
\end{align}
where $F_{\gamma l}$ are the multipole moments of the photon temperature perturbation, $k$ is the Fourier wavenumber, $\dot{\kappa} = a n_{\rm e} \sigma_{\rm T}$ is the Thomson opacity, $\delta_{\rm b}$ is the baryon density perturbation, $v_{\rm b}$ is the baryonic bulk velocity, $c_{\rm s}$ is the baryonic sound speed, and $\phi$ and $\psi$ are the two gravitational potentials in conformal Newtonian gauge. Note that we have used the relationship 
\begin{equation}
    \frac{d}{d\eta} = a^2 H \frac{d}{da}
\end{equation}
to convert between conformal time ($\eta$) derivatives and scale-factor derivatives. It is straightforward to see that these equations are invariant under the transformation
\begin{equation}\label{eq:key_transf}
    H\to \lambda H, \, k\to \lambda k,\, \dot{\kappa}\to \lambda \dot{\kappa}.
\end{equation}
These transformations correspond to equally rescaling \emph{all} length scales appearing in the Boltzmann equations: the Hubble horizon, the wavelength of fluctuations, and the photon mean free path.
 To close this system of equations, we need the perturbed Einstein equations for the $\phi$ and $\psi$ potentials. We use here the Poisson and shear equations
\begin{align}\label{eq:Einstein_eq}
    k^2\phi + 3 a H \left(a^2H\frac{d\phi}{da} + a H \psi\right) &= -4\pi G a^2 \sum_i \rho_i \delta_i, \\
    k^2(\phi - \psi) &= 12\pi G a^2 \sum_i (\rho_i + P_i)\sigma_i,\nonumber
\end{align}
where $\delta_i$ and $\sigma_i$ are the energy density perturbation and anisotropic stress of species $i$, respectively. These equations are invariant under the transformation given in Eq.~\eqref{eq:key_transf}, provided that the energy density of each component is individually rescaled, that is, $\sqrt{G\rho_i} \to \lambda \sqrt{G\rho_i}$. Massless neutrinos and dark matter follow collisionless versions of the equation given in Eqs.~\eqref{eq:phot_bar}, implying that they too are invariant under the transformation $H\to \lambda H$ and $k\to \lambda k$. We note that the evolution of massive neutrinos perturbations is also invariant under this transformation, once their masses are also properly scaled.

We thus see that the linear evolution equations of \emph{all} components present in the Universe are invariant under the transformation
\begin{equation}\label{eq:key_transf2}
    \sqrt{G\rho_i} \to \lambda \sqrt{G\rho_i}, \, k\to \lambda k,\, \dot{\kappa}\to \lambda \dot{\kappa}.
\end{equation}
This means that we can express the solution $\tilde{\Phi}$ to the perturbation equations in the presence of this scaling symmetry in terms of the original solution $\Phi$ in the absence of scaling (i.e.~$\lambda =1$) as
\begin{equation}\label{eq:scale_inv}
    \tilde{\Phi}(k,\dot{\kappa},a,\lambda) = \Phi(k/\lambda,\dot{\kappa}/\lambda,a,\lambda=1),
\end{equation}
where $\Phi$ here stands for any of the dimensionless perturbation variables (e.g.~$\delta,v, F_{\gamma l}$, etc.). Such a relation was first presented in Ref.~\cite{Zahn_2003} in the context of the tight-coupling approximation ( $\dot{\kappa} \gg H$), but we see here that it applies in a broader context once the Thomson opacity is also rescaled.

\subsection{Second-order equations}
Equations \eqref{eq:phot_bar} and \eqref{eq:Einstein_eq} are invariant under the scaling transformation given in Eq.~\eqref{eq:key_transf2} essentially on dimensional ground. By consistency, all terms in a single differential equation need to have the same mass (or length) dimensions. Since the scaling transformation uniformly rescale \emph{all} length scales in the problem, then these equations are invariant as all terms are multiplied by the same scaling factor. The same logic applies to higher-order equations. For instance, the second-order equations for photons, baryons, cold dark matter and neutrinos are presented in Ref.~\cite{Bartolo:2007ax}. As in the first-order case, all terms in the relevant differential equations (when written with the scale factor $a$ as the time variable) either involve a factor of $H$, a factor of $k$, or a factor of $\dot{\kappa}$, each multiplied by dimensionless perturbation variables. Such equations are thus invariant under the scaling transformation given in Eq.~\eqref{eq:key_transf2}.

\end{document}